\documentclass[reprint,prl,superscriptaddress,floatfix]{revtex4-2}

\usepackage[table,dvipsnames]{xcolor} % Combined xcolor with table option; removed redundant 'color' package
\usepackage{graphicx}

\usepackage{amsmath}
\usepackage{amsfonts}
\usepackage{amssymb}
\usepackage{bm}

\usepackage{siunitx}
\usepackage{framed}
\usepackage{adjustbox}
\usepackage{tabularx}
\usepackage{booktabs}
\usepackage{float}
\usepackage{makecell}
\usepackage[normalem]{ulem}

\usepackage{hyperref}
\hypersetup{
    colorlinks=true,
    linkcolor=blue,
    filecolor=magenta,
    urlcolor=cyan,
}

\providecommand{\U}[1]{\protect\rule{.1in}{.1in}}

\makeatletter
\renewcommand*{\fnum@figure}{{\normalfont\bfseries \figurename~\thefigure}}
\providecommand*{\@caption@fignum@sep}{\textbf{ : }}
\renewcommand*{\@caption@fignum@sep}{\textbf{ : }}
\makeatother

\begin{document}

\title{Two Unconventional Hall Effects in the Kagome Magnet TmMn$_4$Cr$_2$Sn$_6$}

%Imposter topological effects
% -------------------------------------------------------------------

\author{Mohamed El Gazzah}
\email{melgazza@nd.edu}
\thanks{Corresponding Author}
\affiliation{Department of Physics and Astronomy, University of Notre Dame, Notre Dame, IN 46556, USA}
\affiliation{Stavropoulos Center for Complex Quantum Matter, University of Notre Dame, Notre Dame, IN 46556, USA}
\affiliation{Center for Nanophase Materials Sciences, Oak Ridge National Laboratory, Oak Ridge, TN 37830, USA}

\author{Fehmi Sami Yasin}
\email{yasinfs@ornl.gov}
\thanks{Corresponding Author}
\affiliation{Center for Nanophase Materials Sciences, Oak Ridge National Laboratory, Oak Ridge, TN 37830, USA}

\author{Bishal Thapa}
\affiliation{Department of Physics and Astronomy, George Mason University, Fairfax, VA 22030, USA}
\affiliation{Quantum Science and Engineering Center, George Mason University, Fairfax, VA 22030, USA}

\author{Sk Jamaluddin}
\affiliation{Department of Physics and Astronomy, University of Notre Dame, Notre Dame, IN 46556, USA}
\affiliation{Stavropoulos Center for Complex Quantum Matter, University of Notre Dame, Notre Dame, IN 46556, USA}

\author{Daniel Lu}
\affiliation{Department of Physics and Astronomy, University of Notre Dame, Notre Dame, IN 46556, USA}
\affiliation{Stavropoulos Center for Complex Quantum Matter, University of Notre Dame, Notre Dame, IN 46556, USA}

\author{Abhijeet Nayak}
\affiliation{Department of Physics and Astronomy, University of Notre Dame, Notre Dame, IN 46556, USA}
\affiliation{Stavropoulos Center for Complex Quantum Matter, University of Notre Dame, Notre Dame, IN 46556, USA}

\author{Resham Regmi}
\affiliation{Department of Physics and Astronomy, University of Notre Dame, Notre Dame, IN 46556, USA}
\affiliation{Stavropoulos Center for Complex Quantum Matter, University of Notre Dame, Notre Dame, IN 46556, USA}

\author{Mary Katherine Carroll}
\affiliation{Department of Physics and Astronomy, University of Notre Dame, Notre Dame, IN 46556, USA}
\affiliation{Stavropoulos Center for Complex Quantum Matter, University of Notre Dame, Notre Dame, IN 46556, USA}

\author{Igor I. Mazin}
\affiliation{Department of Physics and Astronomy, George Mason University, Fairfax, VA 22030, USA}
\affiliation{Quantum Science and Engineering Center, George Mason University, Fairfax, VA 22030, USA}

\author{Nirmal J. Ghimire}
\email{nghimire@nd.edu}
\thanks{Corresponding Author}
\affiliation{Department of Physics and Astronomy, University of Notre Dame, Notre Dame, IN 46556, USA}
\affiliation{Stavropoulos Center for Complex Quantum Matter, University of Notre Dame, Notre Dame, IN 46556, USA}
\date{\today}
%*********************************************************************************************************************************

\begin{abstract}
We report two Hall anomalies in the kagome magnet TmMn$_4$Cr$_2$Sn$_6$. A high-field anomaly is attributed to a discontinuous spin reorientation driven by non-monotonic Tm anisotropy and differing sublattice contributions to Hall transport and magnetization. At low fields, Hall hysteresis persists despite negligible magnetic hysteresis, consistent with a chirality-related contribution whose microscopic origin remains unresolved. These results distinguish a transition-induced Hall residual from a zero-field response beyond conventional magnetization scaling.
\end{abstract}

\maketitle

% -------------------------------------------------------------------

% -------------------------------------------------------------------

The interplay between magnetic order and electronic transport in quantum materials~\cite{gilbert_topological_2021,chen_anomalous_2014,nagaosa_anomalous_2006} offers a fertile playground for discovering novel topological magnetic states~\cite{luo_skyrmion_2021,noauthor_topological_nodate-2}. In systems where localized magnetic moments couple strongly to itinerant charge carriers, the underlying spin configuration can profoundly influence the electronic wavefunctions, leading to highly unconventional transport signatures. A particularly powerful manifestation of this coupling is the topological Hall effect (THE)~\cite{neubauer_topological_2009,gobel_topological_2025,taylor_anomalous_2020}, which, if identified correctly, can serve as a sensitive transport signature of topologically nontrivial spin textures in real space. Unlike the ordinary Hall effect (OHE) driven by the Lorentz force, or the anomalous Hall effect (AHE) caused by momentum-space Berry curvature~\cite{nagaosa_anomalous_2006,chen_anomalous_2014,smejkal_emerging_2022}, the THE arises when conduction electrons accumulate a geometric Berry phase as they adiabatically traverse a localized spin texture with finite scalar spin chirality, $\chi_{ijk} = \mathbf{S}_i \cdot (\mathbf{S}_j \times \mathbf{S}_k)$, where $\mathbf{S}_i,\mathbf{S}_j,\mathbf{S}_k$ are three neighboring spins~\cite{noauthor_competing_nodate}. Consequently, observation of a robust THE is often viewed as compelling evidence for the stabilization of complex magnetic topologies~\cite{
nagaosa_topological_2013, tatara_chirality-driven_2002,noauthor_competing_nodate}.

\begin{figure*}
    \centering
    \includegraphics[width=1\linewidth]{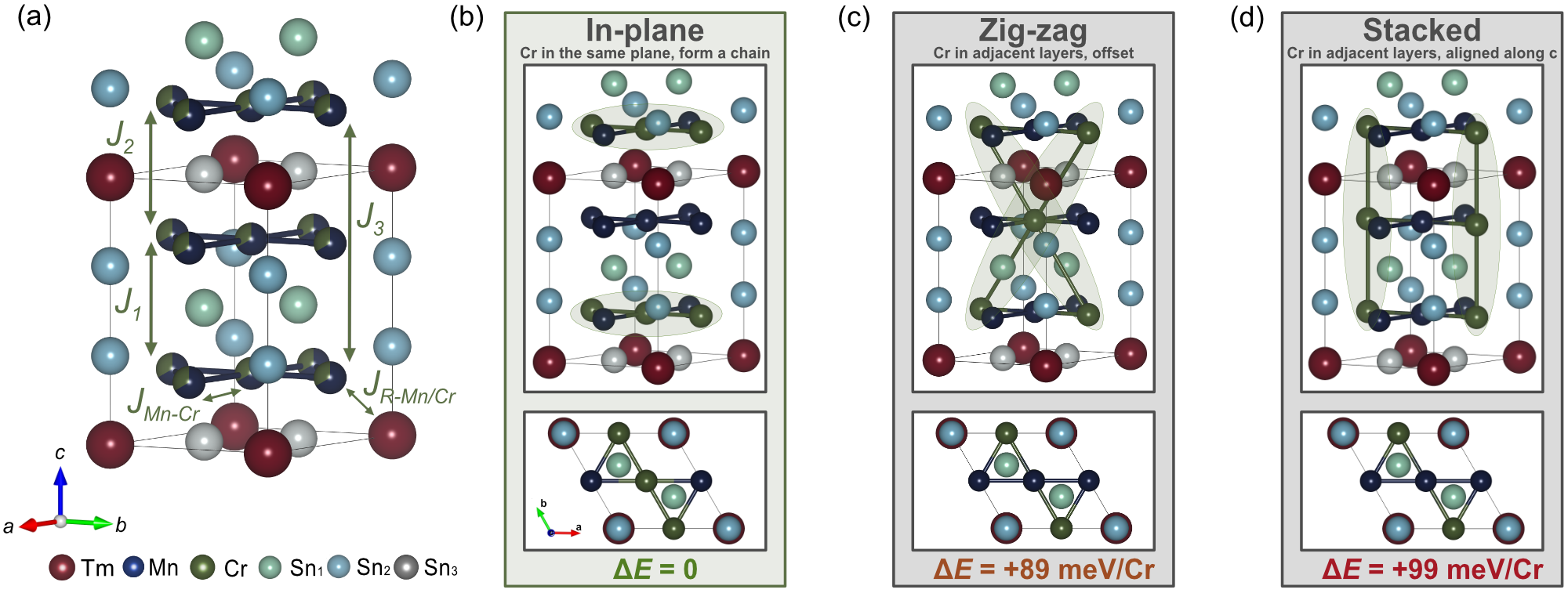}
    \caption{Crystal structure of TmMn$_4$Cr$_2$Sn$_6$ and candidate Cr arrangements within the kagome layers. (a)~Hexagonal unit cell of the parent $R$Mn$_6$Sn$_6$ structure (space group $P6/mmm$, No.~191), showing the Mn/Cr kagome layers separated by Tm and Sn layers. Interlayer exchange interactions $J_1$, $J_2$, and $J_3$ couple the kagome layers along the $c$-axis, and $J_{R-\mathrm{Mn/Cr}}$ denotes the exchange between the Tm--Sn layer and the kagome layers. The nearest-neighbor in-plane exchange interaction within the Mn/Cr layers is denoted $J_{\mathrm{Mn-Cr}}$. (b)--(d)~The three inequivalent ways of distributing two Cr and four Mn atoms over the six kagome sites of the unit cell, each shown in perspective (upper) and projected along [001] (lower), together with the relative total energy per Cr obtained from open-core DFT. (b)~\emph{In-plane}: both Cr occupy the same kagome layer on neighboring sublattices, forming a Cr dimer. (c)~\emph{Zigzag}: Cr occupy adjacent layers on different sublattices, offset in the basal plane. (d)~\emph{Stacked}: Cr occupy adjacent layers on the same sublattice, aligned along $c$.}
    \label{Fig1}
\end{figure*}

The standard Hall decomposition partitions the Hall resistivity into an OHE term proportional to the applied field, an AHE term proportional to the net magnetization, and a residual often labeled as a THE. This treatment is reliable when the ordinary and anomalous Hall coefficients remain approximately unchanged within a magnetic phase and when the anomalous Hall response and the net magnetization are both proportional to the domain population imbalance. These assumptions can fail in magnetically complex systems where applied fields can trigger spin reorientations and metamagnetic transitions that dramatically alter the OHE or AHE, producing a spurious residual signal that mimics a THE. This is the case in YMn$_6$Sn$_6$, where the OHE changes sign between the low-field spiral state and the high-field forced FM state~\cite{noauthor_competing_nodate}, as well as in the spin-ice compound HoAgGe, where a field-driven Lifshitz transition results in a dramatic change in OHE~\cite{bhandari2025tunable}. This underscores the importance of carefully disentangling the OHE, AHE, and THE contributions in order to categorize them accurately.

In this Letter, we report two distinct unconventional Hall-effect (UHE) regimes in TmMn$_4$Cr$_2$Sn$_6$, a centrosymmetric kagome magnet formed by substituting Cr onto the Mn kagome site~\cite{schobinger-papamantellos_magnetic_2000, destefano_giant_2025} of TmMn$_6$Sn$_6$. Hall transport reveals a zero-field UHE at low temperatures ($5~\mathrm{K}<T\lesssim75~\mathrm{K}$) and a separate high-field UHE pocket at intermediate temperatures ($40~\mathrm{K}\lesssim T\lesssim100~\mathrm{K}$). The zero-field UHE is hysteretic and disproportionate to the bulk magnetic hysteresis, consistent with a Hall contribution beyond simple magnetization scaling and plausibly associated with frozen non-coplanar magnetism. The high-field anomaly, by contrast, follows a field-induced magnetic transition and is naturally explained by a breakdown of the conventional magnetization-based Hall decomposition. Thus, TmMn$_4$Cr$_2$Sn$_6$ provides an example in which phenomenologically similar Hall residuals can arise from physically distinct mechanisms.

The parent compound TmMn$_6$Sn$_6$ crystallizes in the $P6/mmm$ structure (space group No.~191), as shown in Fig.~\ref{Fig1}(a), with key exchange interactions $J_1$, $J_2$, $J_3$, $J_{\mathrm{Mn-Cr}}$, and $J_{R-\mathrm{Mn/Cr}}$ driving the overall magnetism. Partial substitution of Mn with Cr can thus lead to both structural (such as reduction of symmetry) and magnetic changes. Refinement of the single-crystal X-ray diffraction (XRD) data (Table~\ref{T1} of the Supplemental Material (SM)) shows that Cr is distributed over the 6$i$ site of both kagome layers. A preferential layer-by-layer arrangement, in which Cr-rich and Mn-rich kagome planes alternate, would render the two kagome layers inequivalent and lower the crystal symmetry to $P6mm$ (No.~183). The diffraction data, however, index fully in $P6/mmm$. Because the X-ray scattering contrast between Cr and Mn amounts to a single electron, these data constrain long-range Cr order but not short-range correlations.

To gain further insight into the preferred Cr arrangement and its magnetic consequences, we performed open-core DFT calculations for the possible arrangements of two Cr and four Mn ions within a single unit cell, fully relaxing the ionic positions while keeping the unit cell dimensions fixed. Three inequivalent configurations are available (Figs.~\ref{Fig1}(b)--\ref{Fig1}(d)): both Cr within the same kagome layer on neighboring sublattices, forming in-plane dimers (in-plane, Fig.~\ref{Fig1}(b)); Cr distributed between adjacent layers on different sublattices, offset within the basal plane (zigzag, Fig.~\ref{Fig1}(c)); and Cr in adjacent layers on the same sublattice, aligned along $c$ (stacked, Fig.~\ref{Fig1}(d)).

We find that Cr preferentially forms dimers within a single kagome layer despite the steric penalty implied by the Cr--Mn ionic radius difference. These dimers are antiferromagnetically (AFM) ordered, leaving the Cr subsystem magnetically compensated despite large local moments of $M \approx 2.8\ \mu_{\mathrm{B}}$ per Cr site, consistent with a high-spin Cr$^{3+}$ configuration. This agrees with earlier calculations for Tb(Mn$_{1-x}$Cr$_x$)$_6$Sn$_6$~\cite{destefano_giant_2025}.

For the configuration with alternating Cr$_2$Mn and Mn$_3$ kagome planes, we evaluated the effective anisotropy of the Cr dimers by rotating their moments out of the basal plane while keeping the Mn moments in-plane. The calculated energy cost is 10.7 (11.2)~meV/Cr in the open-core approximation with (without) SOC. Rotating the Cr moments by $90^\circ$ within the basal plane incurs nearly the same energy cost as an equivalent out-of-plane rotation, with a difference of less than 0.15~meV/Cr.

Since the AFM Cr dimers are decoupled from the FM Mn moments at the bilinear (Heisenberg) level, and SOC has little effect on the rotation energy, these results indicate a Cr--Mn coupling beyond the Heisenberg model. We describe this coupling by a biquadratic term with a negative coupling constant of order 10~meV/Cr. Including the Tm 4$f$ states increases the energy difference to 16.2~meV/Cr, suggesting an additional, weaker biquadratic interaction with Tm. Finally, forcing the Cr moments into FM alignment drives them into a low-spin state ($M \approx 0.63\ \mu_{\mathrm{B}}$ per Cr site) that is energetically highly unfavorable relative to the AFM configuration.
\begin{figure}
    \centering
    \includegraphics[width=1\linewidth]{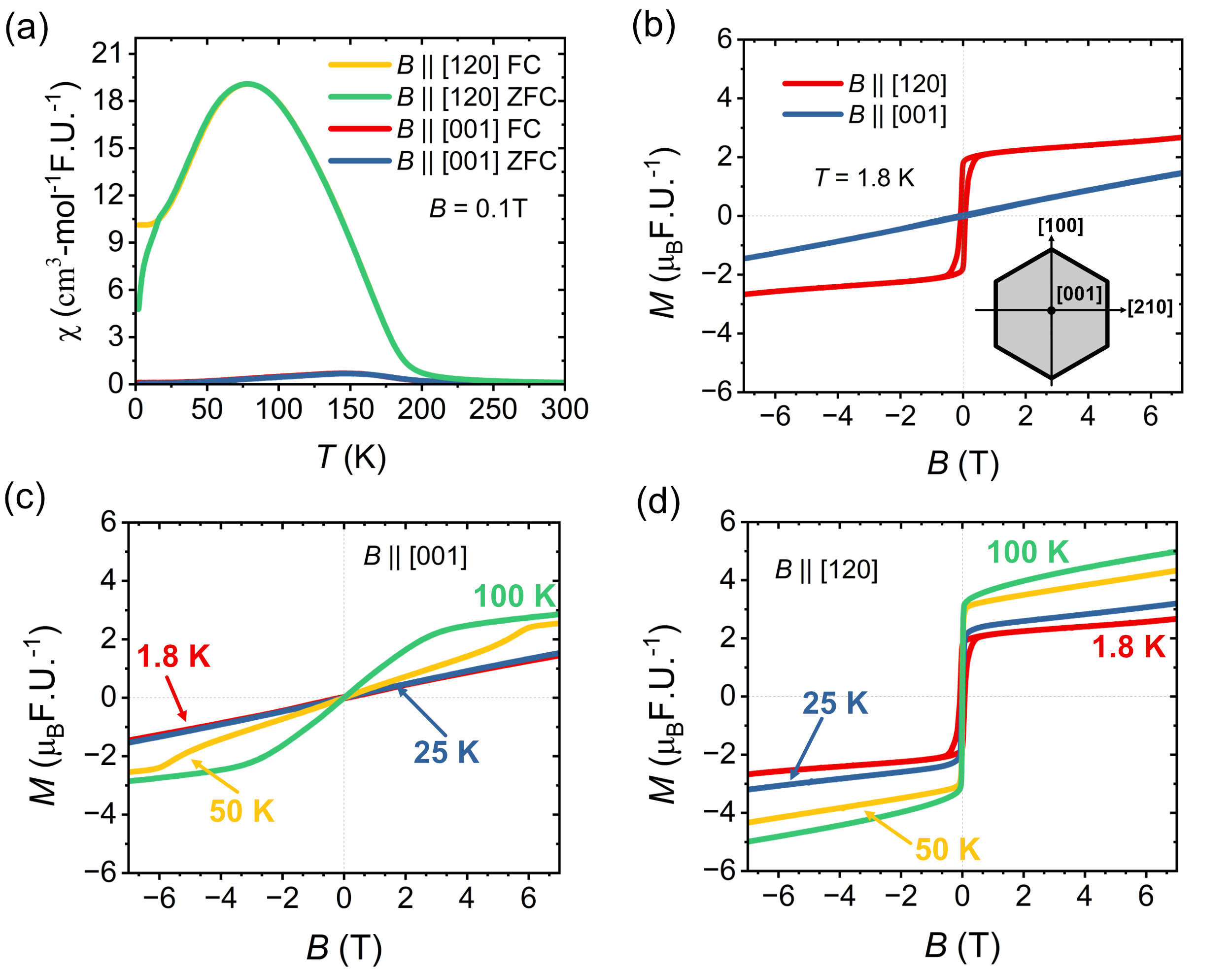}
    \caption{Magnetic characterization of TmMn$_4$Cr$_2$Sn$_6$. (a) Temperature-dependent susceptibility $\chi(T)$ measured at $B = 0.1$~T under field-cooled (FC) and zero-field-cooled (ZFC) conditions for $B \parallel [120]$ and $B \parallel [001]$. (b) Magnetization $M(B)$ at $T = 1.8$~K for $B \parallel [120]$ and $B \parallel [001]$. (c),(d) $M(B)$ at selected temperatures between 1.8 and 100~K for $B \parallel [001]$ and $B \parallel [120]$, respectively.}
    \label{Fig2}
\end{figure}

Figure~\ref{Fig2}(a) shows the temperature-dependent magnetic susceptibility measured under zero-field-cooled (ZFC) and field-cooled (FC) protocols in an applied field of $B=\mu_{0}H=0.1$~T along [120] and [001]. The susceptibility along [120] exhibits a broad maximum near $80$~K, whereas that along [001] is more than an order of magnitude smaller and only weakly temperature dependent. At $1.8$~K (Fig.~\ref{Fig2}(b)), the magnetization along [001] is nearly linear, while the in-plane magnetization saturates rapidly near $300$~mT, indicating easy-plane anisotropy. A small hysteresis is nevertheless resolved along [001] at $1.8$~K (Fig.~\ref{FigSHysteresis} of the SM). With increasing temperature (Figs.~\ref{Fig2}(c)--\ref{Fig2}(d)), the hysteresis along [120] decreases, while a cusp appears in the [001] magnetization at $50$~K, marking the onset of the forced FM phase.

\begin{figure}
    \centering
    \includegraphics[width=1\linewidth]{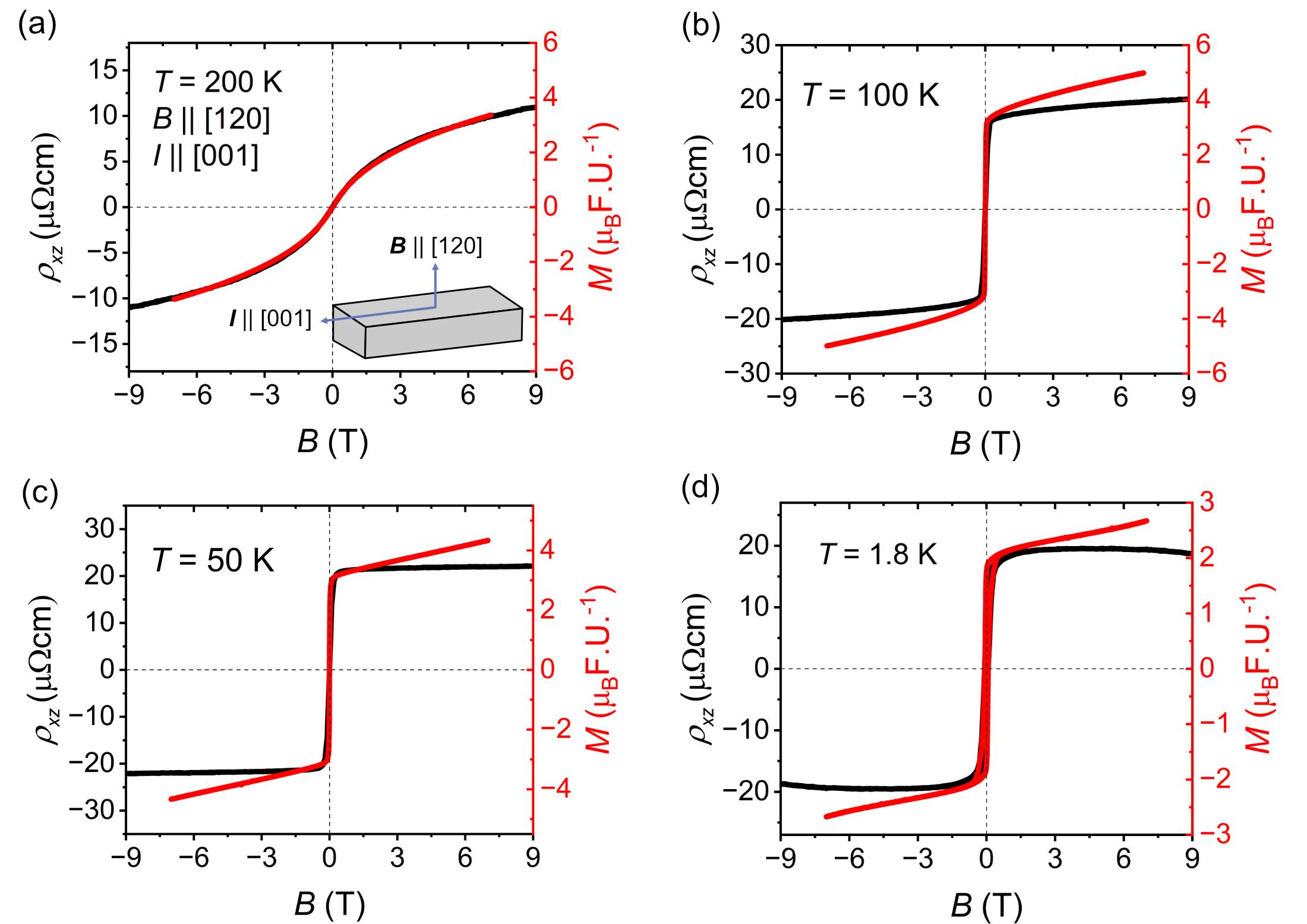}
    \caption{Hall resistivity and magnetization for $B \parallel [120]$ and $I \parallel [001]$. Field-dependent Hall resistivity $\rho_{xz}(B)$ of TmMn$_4$Cr$_2$Sn$_6$, overlaid with the corresponding magnetization $M(B)$ plotted on the right axis. Data are shown at (a) 200 K, (b) 100 K, (c) 50 K, and (d) 1.8 K.}
    \label{Fig3}
\end{figure}

This behavior supports a model in which AFM-coupled Tm and Mn moments order ferrimagnetically (FiM). The in-plane anisotropy of the hexagonal Mn plane is expected to be small, arising only at sixth order in (already weak) Mn$^{2+}$ spin-orbit coupling. The nearly hysteresis-free in-plane magnetization above the lowest temperatures (e.g., at $50$~K for $B \parallel [120]$, Fig.~\ref{Fig3}(c)) suggests that the Tm in-plane anisotropy is also small. The measured low-temperature, low-field magnetization of $\sim 2~\mu_{\mathrm{B}}$ per formula unit agrees reasonably with the DFT estimate: subtracting the antiparallel Tm $f$-shell moment of $7~\mu_{\mathrm{B}}$ from the open-core result of $\sim 9.4~\mu_{\mathrm{B}}$ yields $\sim 2.4~\mu_{\mathrm{B}}$ per formula unit, with essentially no Cr contribution at base temperature.

At higher temperatures or fields, the Cr dimers begin to contribute to the magnetization. This contribution is reflected in the increasing low-field moment ($\sim 2$, $3.5$, and $4~\mu_{\mathrm{B}}$ per formula unit at $1.8$, $50$, and $100$~K, respectively) and an additional field-linear susceptibility of approximately $0.1$--$0.2~\mu_{\mathrm{B}}$/T per formula unit above the constant FiM background (Figs.~\ref{Fig2}(d) and \ref{Fig3}).

\begin{figure*}
    \centering
    \includegraphics[width=0.8\linewidth]{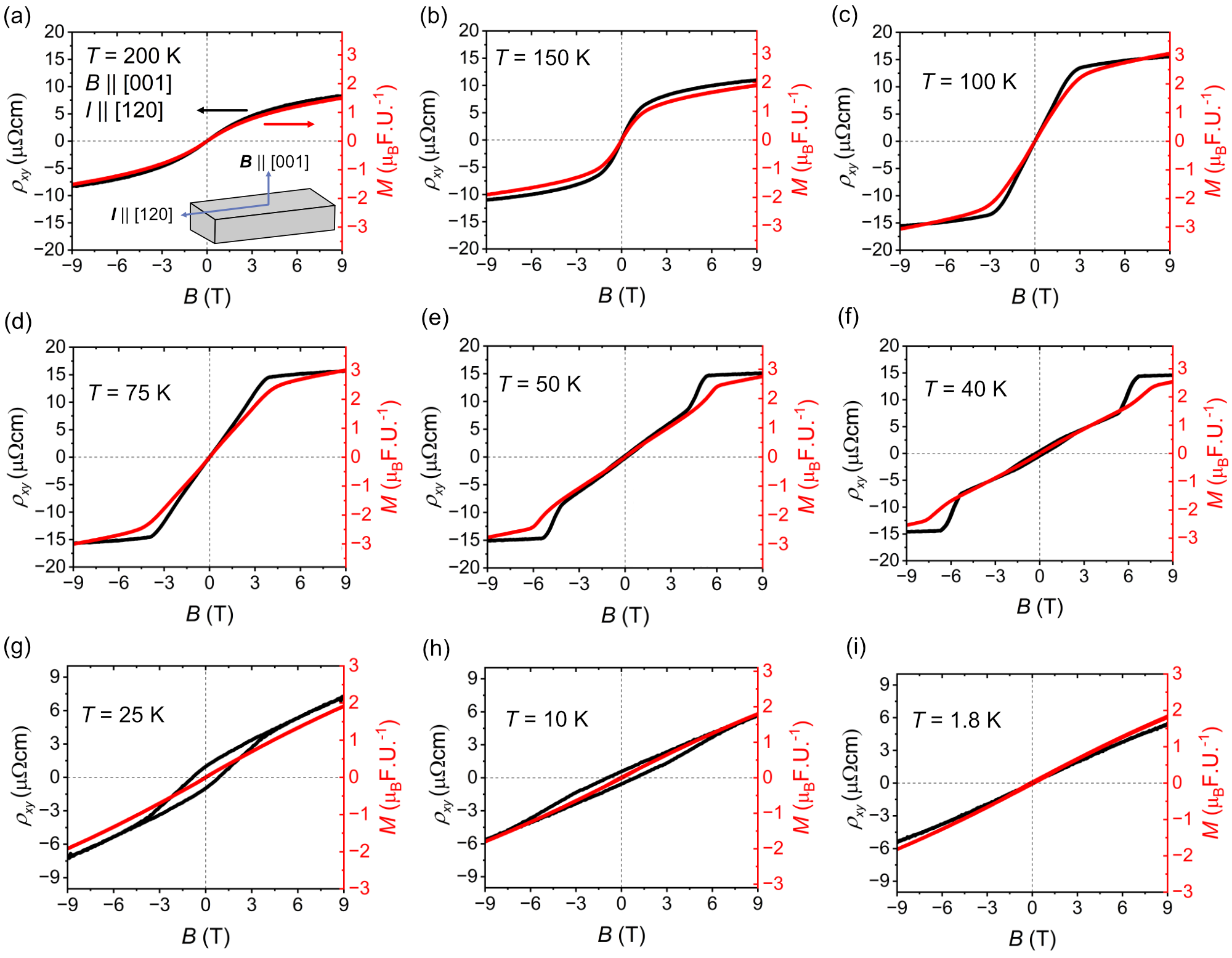}
    \caption{Hall resistivity and magnetization for $B \parallel [001]$ and $I \parallel [120]$.
Field-dependent Hall resistivity $\rho_{xy}$ of TmMn$_4$Cr$_2$Sn$_6$, overlaid with the corresponding magnetization $M(B)$ plotted on the right axis. Data are shown at (a) 200 K, (b) 150 K, (c) 100 K, (d) 75 K, (e) 50 K, (f) 40 K, (g) 25 K, (h) 10 K, and (i) 1.8 K. At high temperatures, $\rho_{xy}$ largely follows the smooth field evolution of $M(B)$. Upon cooling below $\sim 100$ K, the Hall response develops pronounced field-induced anomalies that are no longer proportional to the bulk magnetization.} 
    \label{Fig4}
\end{figure*}

We begin our analysis of the Hall transport by examining the in-plane field geometry, where the magnetic field $B \parallel [120]$ and the current $I \parallel [001]$, with the Hall resistivity denoted $\rho_{xz}$. The field-dependent evolution of $\rho_{xz}$ (black curves) is presented in Fig.~\ref{Fig3} alongside the corresponding magnetization loops (red curves) for temperatures from $200\text{ K}$ down to $1.8\text{ K}$. At $200\text{ K}$ (Fig.~\ref{Fig3}(a)), $\rho_{xz}(B)$ and $M(B)$ have the same  field dependence, tracking one another closely through saturation. Upon cooling to $100\text{ K}$ and below (Figs.~\ref{Fig3}(b)--\ref{Fig3}(d)), $\rho_{xz}$ closely tracks $M$ at low fields; however, at higher fields, $\rho_{xz}$ saturates abruptly while $M$ continues to increase nearly linearly.
As discussed above, this high-field increase in $M$, as well as its distinct temperature dependence, is attributable to canting within the Cr dimers. The Hall resistivity is insensitive to this canting: $\rho_{xz}$ remains practically constant and temperature-independent once the FiM state is established. This confirms that the Hall transport in the in-plane geometry is determined nearly exclusively by the Mn/Tm sublattice magnetization, with a negligible Cr contribution.

We now turn to the out-of-plane field geometry, which reveals a distinctly richer behavior. With $B \parallel [001]$ and $I \parallel [120]$, the Hall resistivity $\rho_{xy}$ displayed in Figs.~\ref{Fig4}(a)--\ref{Fig4}(i) is overlaid with the corresponding magnetization curves (red). At high temperatures ($200\text{ K}$), $\rho_{xy}$ closely follows the trend of $M(B)$. As the temperature is lowered to $150\text{ K}$ and $100\text{ K}$, $\rho_{xy}$ begins to deviate from the bulk magnetization, saturating slightly before $M$. This divergence becomes pronounced at $75\text{ K}$, where $\rho_{xy}$ exhibits a sharp saturation profile while $M$ displays a much smoother, more gradual transition.

This feature is far clearer at $50$ and $40\text{ K}$ (Figs.~\ref{Fig4}(e)--\ref{Fig4}(f)), where a sharp high-field Hall jump yields a much stronger relative signature than the subtle kink in $M(B)$. Simultaneously, a low-field Hall hysteresis appears, yielding finite remanent $\rho_{xy}$ despite only weak magnetic hysteresis. Below $40\text{ K}$, the spin reorientation shifts beyond the available $9\text{ T}$ field window (Figs.~\ref{Fig4}(g)--\ref{Fig4}(i)) while the hysteretic response in the Hall channel is amplified.

\begin{figure}
    \centering
    \includegraphics[width=1\linewidth]{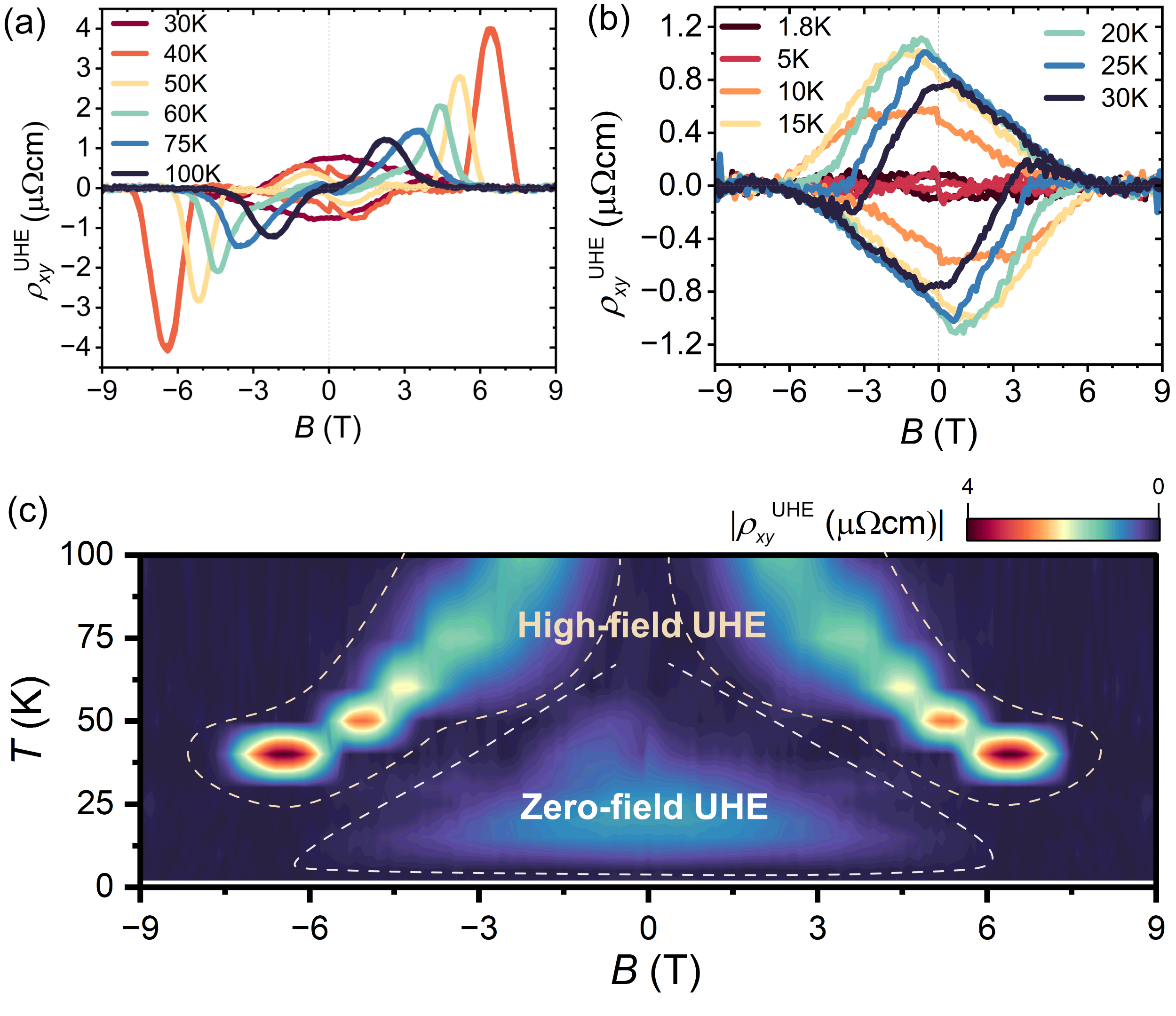}
    \caption{Extracted unconventional Hall contribution in TmMn$_4$Cr$_2$Sn$_6$ for $B \parallel [001]$ and $I \parallel [120]$. $\rho_{xy}^{\mathrm{UHE}}$ is obtained from the data in Fig.~\ref{Fig4} using Eq.~(\ref{eqHall1}) and the fitting protocol in Sec.~\ref{sec:hall} of the SM. (a) High-field peaks associated with the spin-reorientation transition at 40--100~K. (b) Zero-field hysteretic response at 1.8--30~K, with both sweep directions shown. The 30~K curve, which shows no high-field UHE, appears in both panels for comparison of the high-field and zero-field responses. (c) Interpolated map of $|\rho_{xy}^{\mathrm{UHE}}|$ in the $B$--$T$ plane; dashed contours outline the high-field and zero-field regimes.} 
    \label{Fig5}
\end{figure}

To systematically extract the transport-magnetization deviations, the total out-of-plane Hall response is conventionally decomposed as~\cite{noauthor_competing_nodate}
\begin{equation}
    \rho_{xy} = \rho_0 + \rho_A + \rho_{xy}^{\mathrm{UHE}} = R_0B + R_s\mu_0M + \rho_{xy}^{\mathrm{UHE}},
    \label{eqHall1}
\end{equation}
where $\rho_0 = R_0B$ is the ordinary linear Hall resistivity dictated by the effective carrier density, $\rho_A = R_s\mu_0M$ is a phenomenological magnetization-linked anomalous Hall contribution, and $\rho_{xy}^{\mathrm{UHE}}$ is the residual that quantifies deviations from the assumed scaling. We emphasize that $\rho_{xy}^{\mathrm{UHE}}$ is operationally defined and is not, by itself, equivalent to a topological Hall resistivity. The background fitting and subtraction protocol is detailed in Sec.~\ref{sec:hall} of the SM. The isolated $\rho_{xy}^{\mathrm{UHE}}(B)$ profiles, compiled in Fig.~\ref{Fig5}, reveal that the observed transport-magnetization decoupling manifests as two distinct anomalies well separated in the $B$--$T$ plane.

\begin{figure}
    \centering
    \includegraphics[width=0.8\linewidth]{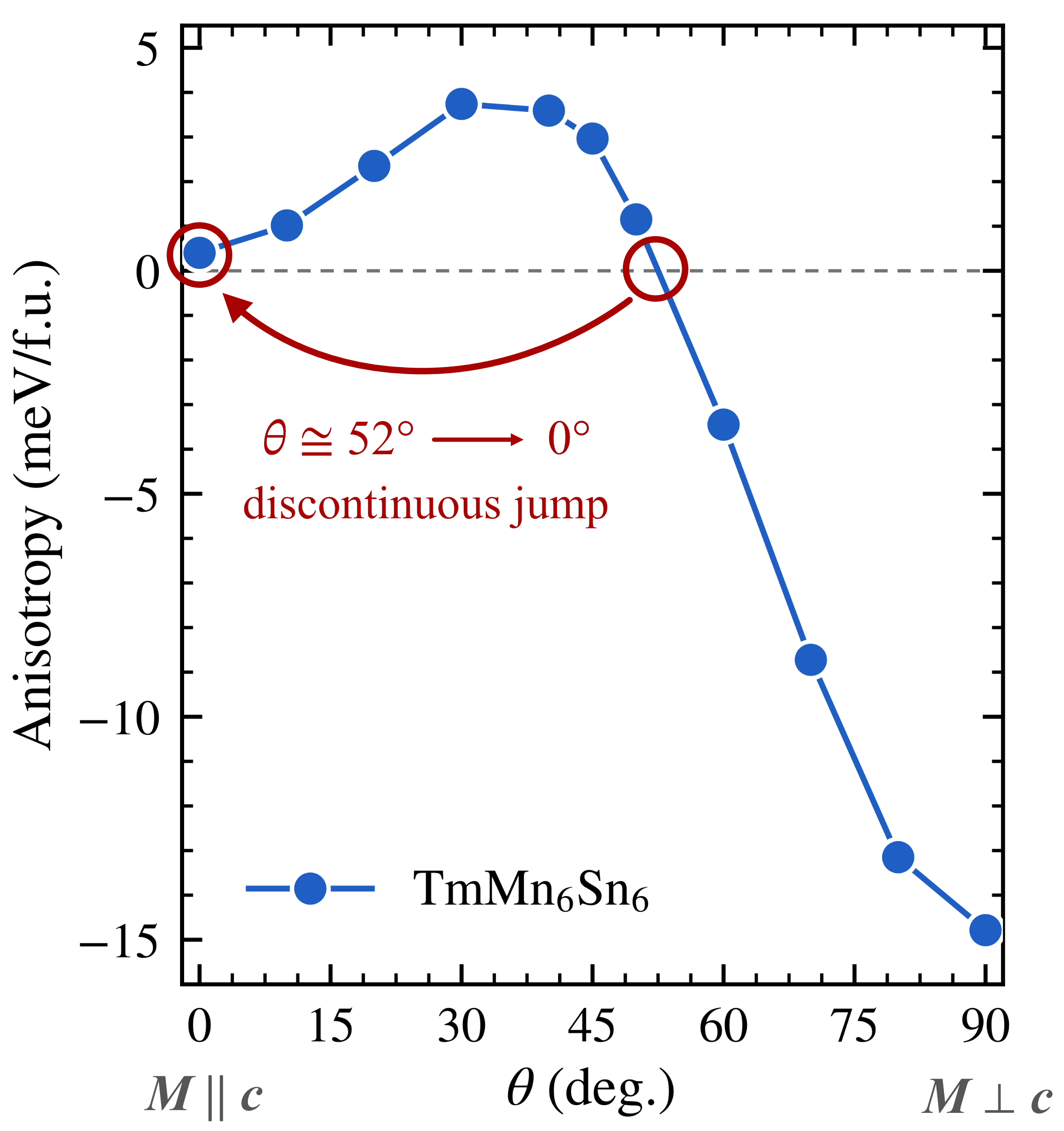}
    \caption{Calculated magnetic anisotropy energy of TmMn$_6$Sn$_6$ as a function of the polar angle $\theta$ between the Tm moment and the $c$-axis. The energy is non-monotonic, exhibiting a local maximum near $\theta \approx 30^\circ$ and crossing zero near $\theta \approx 52^\circ$ before decreasing steeply toward the easy-plane minimum at $\theta = 90^\circ$. The red arrow marks the discontinuous jump from $\theta \approx 52^\circ$ to $\theta = 0^\circ$ discussed in the text. Data adapted with permission from M.~El~Gazzah \textit{et al.}, Phys.\ Rev.\ Materials \textbf{9}, 114414 (2025)~\cite{noauthor_doping-induced_nodate}, \href{https://doi.org/10.1103/qbwt-qdt7}{doi:10.1103/qbwt-qdt7}. Copyright 2025 by the American Physical Society.}
    \label{Anisotropy}
\end{figure}

The high-field $\rho_{xy}^{\mathrm{UHE}}$ anomaly appears at $100$~K and increases rapidly upon cooling, reaching its maximum of $4.1~\mu\Omega\,\mathrm{cm}$ at $40$~K (Fig.~\ref{Fig5}(a)). Below $40$~K, the anomaly shifts beyond our $9$~T window. Although this residual could be interpreted as a real-space spin-chirality contribution, we propose that it arises from applying an invalid decomposition to a magnetically complex system. Calculations for the parent compound TmMn$_6$Sn$_6$ reveal a non-monotonic Tm anisotropy energy with a local maximum near $\theta\approx30^\circ$ (Fig.~\ref{Anisotropy})~\cite{noauthor_doping-induced_nodate}. Its magnitude, approximately $15$~meV per formula unit, would imply a saturation field near $90$~T and a susceptibility of only $\sim0.05~\mu_{\mathrm{B}}$/T. Both differ substantially from the measured values, indicating a reduced effective anisotropy in the Cr-substituted compound. 
Still, the average crystal field around Tm should be qualitatively the same as in TmMn$_6$Sn$_6$, even if weaker, thus strongly suggesting a similar non-monotonic behavior.
%We nevertheless propose that the qualitative effect of the non-monotonic profile persists.

An immediate ramification of that is a discontinuous spin reorientation rather than smooth canting toward the out-of-plane direction. In the parent-compound model, the moments jump from $\theta\approx52^\circ$ to $\theta=0^\circ$ at a critical spin-reorientation field $B_{\mathrm{sr}}$. A similar transition would explain the sharp Hall jump preceding saturation in Figs.~\ref{Fig4}(e)--\ref{Fig4}(f). The AHE is dominated by the Mn/Tm moments and is largely insensitive to Cr, so $\rho_{xy}$ primarily registers the sharp Mn/Tm spin reorientation. Bulk magnetization instead combines this transition with the smooth, field-linear canting of the weakly coupled Cr moments, producing a more gradual approach to saturation. This difference in sublattice sensitivity explains the mismatch between the Hall and magnetization profiles and the large residual obtained from conventional AHE subtraction.
Below the spin-reorientation transition, an intermediate field window is recovered in which conventional Hall scaling holds. At $40$~K, this window extends from approximately $2.5$ to $5$~T, where $\rho_{xy}$ tracks $M(B)$ and the extracted $\rho_{xy}^{\mathrm{UHE}}$ is negligible. This behavior is consistent with a canted FiM state whose Hall response is described by ordinary and magnetization-linked anomalous contributions.

At still lower fields, a distinct hysteretic Hall contribution appears around zero field despite negligible magnetic hysteresis (Fig.~\ref{Fig5}(b)). Upon cooling below $75$~K, the extracted remanent $\rho_{xy}^{\mathrm{UHE}}$, i.e., the zero-field residual remaining after subtraction of the ordinary and anomalous contributions in Eq.~(\ref{eqHall1}), reaches a maximum of $1.13~\mu\Omega\,\mathrm{cm}$ at $20$~K before becoming negligible below $5$~K (Fig.~\ref{Fig5}(b)). The loop broadens on cooling, with increasing coercivity and no resolved switching within $9$~T at $1.8$~K (Figs.~\ref{Fig4}(f)--\ref{Fig4}(i) and \ref{Fig5}(c)). This decoupling from bulk magnetization suggests a contribution beyond conventional AHE scaling. We propose that Cr moments freeze into a non-coplanar configuration with finite scalar spin chirality but small net magnetization, analogous to the proposed origin of the spontaneous Hall response in $\mathrm{CoNb_3S_6}$~\cite{ghimire_large_2018,takagi_spontaneous_2023,kirstein_topological_2025}.

To investigate this possibility, we performed cryo-LTEM~\cite{Kim2026Ultramicroscopy} imaging on $ab$- and $ac$-plane lamellae down to $4.6$~K and up to $500$~mT (Sec.~\ref{sec:ltem} of the SM). Imaging revealed saw-tooth and striped domains and a small circular spin texture reminiscent of a type-II skyrmion bubble, all hundreds of nanometers or larger, but no resolved chiral texture. These observations leave open a sub-resolution or magnetically compensated configuration without directly establishing the proposed Cr spin texture.

Together, these results distinguish a high-field Hall anomaly plausibly arising from a discontinuous magnetic reorientation from a zero-field hysteretic contribution consistent with a non-coplanar magnetic configuration, separated by an intermediate regime of conventional Hall scaling. Although the microscopic structures remain unresolved, these contrasting responses emphasize the need for independent constraints on magnetic structure and Hall backgrounds before assigning Hall residuals to a topological origin.

\textit{Acknowledgments}---
N.J.G. acknowledges the support of the National Science Foundation CAREER award DMR-2343536. The work at George Mason University was supported by the National Science Foundation under Award No. DMR-2403804. This work was supported by the U.S. Department of Energy (DOE), Office of Science, Basic Energy Sciences (BES), Materials Sciences and Engineering Division (cryogenic-LTEM). This research used resources at the Center for Nanophase Materials Sciences, a DOE Office of Science User Facility operated by the Oak Ridge National Laboratory. The user proposal was allocated to the Thermo Fisher Scientific Helios 5 CX DualBeam FIB and the FEI Nova 200 Dual-Beam SEM/FIB for sample fabrication, as well as the JEOL NEOARM for imaging, under proposal number CNMS2025-R-03410. This material is based upon work supported by the U.S. Department of Energy, Office of Science, Office of Workforce Development for Teachers and Scientists, Office of Science Graduate Student Research (SCGSR) program. The SCGSR program is administered by the Oak Ridge Institute for Science and Education for the DOE under contract number DE-SC0014664.

\section*{Author contributions}
N.J.G. conceived and coordinated the project. M.E.G. grew the crystals and performed the magnetic and magnetotransport measurements and the analysis, and prepared the lamellae for cryo-LTEM imaging. F.S.Y. and M.E.G. performed the cryo-LTEM imaging. B.T. and I.I.M. carried out first-principles calculations and theoretical modeling. M.E.G., F.S.Y., I.I.M., and N.J.G. wrote the manuscript. All authors contributed to the discussion of the results. All authors have read and approved the manuscript.

\section*{References}\label{sec:4}
\bibliography{ref}

% -----------------------------

% =====================================================================
%                     SUPPLEMENTAL MATERIAL
% =====================================================================
\clearpage
\onecolumngrid
\setcounter{figure}{0}
\setcounter{table}{0}
\setcounter{equation}{0}
\setcounter{section}{0}
\setcounter{subsection}{0}
\renewcommand{\thefigure}{S\arabic{figure}}
\renewcommand{\thetable}{S\arabic{table}}
\renewcommand{\theequation}{S\arabic{equation}}
\renewcommand{\thesection}{S\arabic{section}}
\renewcommand{\theHfigure}{S\arabic{figure}}
\renewcommand{\theHtable}{S\arabic{table}}
\renewcommand{\theHequation}{S\arabic{equation}}
\renewcommand{\theHsection}{S\arabic{section}}

\begin{center}
{\large\bfseries Supplemental Material:\\[2pt] Two Unconventional Hall Effects in the Kagome Magnet TmMn$_4$Cr$_2$Sn$_6$}
\end{center}
\vspace{1em}
\section{Crystal growth and structural characterization} 

Single crystals of TmMn$_4$Cr$_2$Sn$_6$ were grown by the self-flux method~\cite{noauthor_metal_nodate,noauthor_doping-induced_nodate} using Sn as a flux. Tm pieces (Alfa Aesar; 99.9\%), Mn pieces (Alfa Aesar; 99.95\%), Cr powder (Alfa Aesar; 99.95\%), and Sn shots (Alfa Aesar; 99.999\%) were added into a 2~mL aluminum oxide crucible in a molar ratio of 1:4:2:20 and sealed in a fused silica ampule under vacuum. The sealed ampule was heated to 1150 $^{\circ}$C over 10 hours, kept at 1150 $^{\circ}$C for 10 hours, and then cooled to 600 $^{\circ}$C at the rate of 5 $^{\circ}$C/h. Once the furnace reached 600 $^{\circ}$C, the tube was centrifuged to separate the crystals in the crucible from the molten flux. Several well-faceted hexagonal crystals up to 100 mg were obtained in the crucible.

\begin{figure}
    \centering
    \includegraphics[width=1\linewidth]{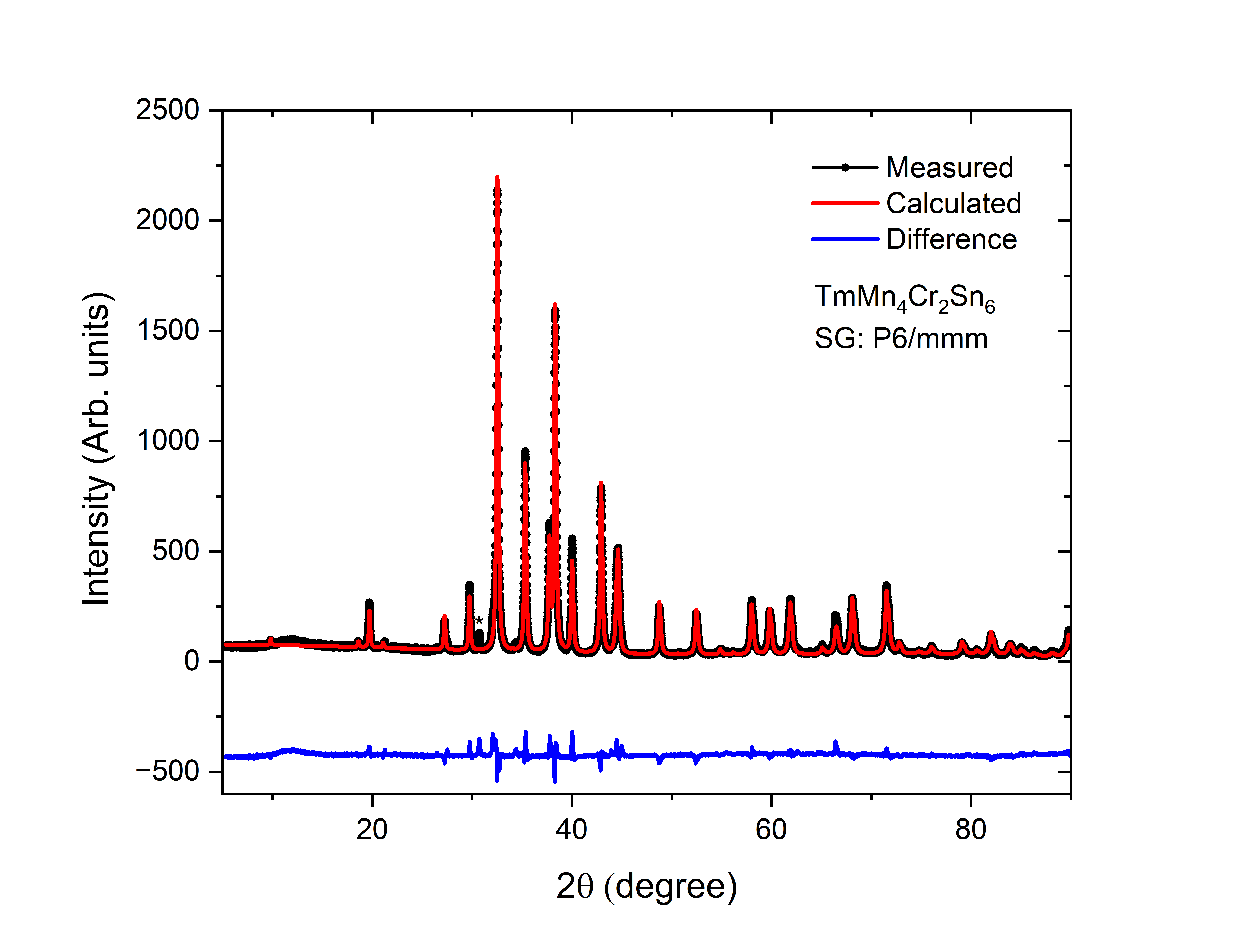}
    \caption{Powder X-ray diffraction refinement of a TmMn$_4$Cr$_2$Sn$_6$ single crystal using the FullProf suite software. The black dotted line shows the measured data, the red curve shows the fitted curve, while the blue curve captures the difference between the two. One impurity peak is highlighted using an asterisk at roughly $2\theta=30^\circ$. This confirms the correct $P6/mmm$ space group is retained despite the Cr substitution.}
    \label{FigS1}
\end{figure}

The crystal structure was first verified by room-temperature powder X-ray diffraction on a Rigaku MiniFlex benchtop diffractometer. A single crystal was cleaned and ground up to a powder, on which the diffraction pattern was collected. The Rietveld refinement performed with the FullProf software is shown in Fig.~\ref{FigS1}, and the resulting lattice parameters and agreement factors are listed in Table~\ref{T2}.

Single-crystal X-ray diffraction was performed to determine the crystal structure of TmMn$_4$Cr$_2$Sn$_6$. Data were collected on a silver-colored, block-shaped crystal using a Bruker Quest diffractometer equipped with a Bruker PHOTON III detector, employing a combination of $\omega$- and $\phi$-scans with a step size of 0.5$^\circ$~\cite{apex4}. The diffraction data were corrected for absorption and polarization effects and analyzed to determine the appropriate space group~\cite{krause2015}. Structure solution was carried out using dual-space algorithms~\cite{sheldrick_shelxt_2015}, followed by routine expansion of the initial model. Final structural refinement was performed using full-matrix least-squares analysis against $F^2$ values for all observed reflections. All atoms were refined with anisotropic atomic displacement parameters. The extinction coefficient was also refined. The crystallographic data and refined atomic parameters are summarized in Table~\ref{T1}.

\begin{table}[h]
\caption{Unit cell parameters and Rietveld agreement factors for TmMn$_{3.87}$Cr$_{2.13}$Sn$_{6}$ from powder X-ray diffraction.
}\label{T2} 
\renewcommand{\arraystretch}{1}% added to adjust size
\small % added to adjust size

\begin{tabular}
 [c]{l@{\hspace{2 cm}}l@{\hspace{2 cm}}l}\hline   
Crystal system & Hexagonal \\
Space group & $P6/mmm$  \\

$a$, $b$ (\AA) & 5.5079(92)\\
$c$ (\AA)   & 9.0145(95)      \\

$R_{WP}$&$13.7\%$ \\
$R_B$ & $5.072\%$ \\
$R_F$ & $3.632\%$ \\
\end{tabular}

\end{table}

\begin{table}[!htbp]
\caption{Crystallographic data, atomic coordinates and equivalent displacement parameters for TmMn$_{3.87}$Cr$_{2.13}$Sn$_{6}$ from single-crystal X-ray diffraction.
}\label{T1} 
\renewcommand{\arraystretch}{1}% added to adjust size
\small % added to adjust size
\begin{tabular}[c]{l@{\hspace{0.15cm}}l}\hline
Crystal system & Hexagonal \\
Space group & $P6/mmm$ \\
Temperature (K) & 120 \\
Wavelength (\AA) & 0.71073 \\
$Z$ (formula units) & 1 \\
$2\theta_{\mathrm{min}}$ & 4.5$^\circ$ \\
$2\theta_{\mathrm{max}}$ & 61.0$^\circ$ \\
Formula weight (g/mol) & 1204.64 \\
$a = b$ (\AA) & 5.4870(3) \\
$c$ (\AA) & 8.9859(4) \\
Volume (\AA$^3$) & 234.29(3) \\
$\mu$ (Mo $K_{\alpha}$) (mm$^{-1}$) & 32.4 \\
Data / parameters & 187 / 15 \\
Extinction coefficient & 0.0107 \\
Goodness-of-fit on $F^2$ & 1.263 \\
Final R indices [$I \ge 2\sigma(I)$] & $R_1 = 0.0099$ \\
 & $wR_2 = 0.0172$ \\
Largest diff.\ peak \& hole (e\,\AA$^{-3}$) & 1.063 and $-0.791$ \\
\end{tabular}

\par%

\begin{tabular}
[c]{c@{\hspace{0.2cm}}c@{\hspace{0.2cm}}c@{\hspace{0.2cm}}c@{\hspace{0.2cm}}c@{\hspace{0.2cm}}c@{\hspace{0.2cm}}c}
\hline
Atom & Wyck & $x$ & $y$ & $z$ & Occ. & $U_{\mathrm{eq}}$ (\AA$^2$) \\ \hline
Tm    & 1b & 0.00000 & 0.00000 & 0.50000 & 1.000 & 0.0044 \\
Sn(1) & 2d & 0.66667 & 0.33333 & 0.50000 & 1.000 & 0.0040 \\
Sn(2) & 2e & 0.00000 & 0.00000 & 0.83656 & 1.000 & 0.0045 \\
Sn(3) & 2c & 0.66667 & 0.33333 & 0.00000 & 1.000 & 0.0044 \\
Mn(1) & 6i & 0.50000 & 0.00000 & 0.25276 & 0.645 & 0.0047 \\ 
Cr(1) & 6i & 0.50000 & 0.00000 & 0.25276 & 0.355 & 0.0047 \\
\hline
\end{tabular}
\end{table}

\section{Magnetization and transport measurements} Resistivity and magnetization measurements were carried out using a Quantum Design DynaCool Physical Property Measurement System (PPMS) and a Quantum Design Magnetic Property Measurement System (MPMS), equipped with 9~T and 7~T magnets, respectively. Single crystals of TmMn$_4$Cr$_2$Sn$_6$ were polished to appropriate dimensions for electrical transport measurements. Crystal orientation along the [100], [120], and [001] directions was determined using a Photonic Science X-ray Laue diffractometer to align the magnetic field along the $[120]$- and $[001]$-axes.

Resistivity and Hall effect measurements were carried out using the standard four-probe method. Platinum wires (25~$\mu$m in diameter) were used for electrical contacts, with contact resistances below 10~$\Omega$. The electrical contacts were affixed using Epotek H20E silver epoxy, and a 2.5~mA current was applied during transport measurements. To correct for contact misalignment, the Hall resistivity was antisymmetrized from the positive and negative applied magnetic fields via $\rho_{\text{H}}=[\rho_\text{T}(+B)-\rho_\text{T}(-B)]/2$, where $\rho_\text{T}$ is the resistivity measured via the transverse voltage contacts in the Hall bar geometry~\cite{bhandari2025tunable}.

The magnetic measurement along the $[001]$-axis was then performed on the same Hall bar used for the Hall resistivity measurement to ensure a reliable correlation between transport and magnetization, and to eliminate sample-dependent demagnetization effects. After the contacts were removed in acetone, the sample was mounted on a brass holder with GE varnish. Because the mass of this sample was too small to be measured accurately, a separate measurement was performed on a bulk single crystal of known mass, and the magnetization of the Hall bar was scaled to it. The magnetic measurements along the $[120]$-axis were conducted in the MPMS, where a single crystal was mounted on a quartz holder using GE varnish and allowed to dry before loading.

\section{Lamella preparation} After confirming the desired composition, we prepared electron-transparent lamellae for Lorentz transmission electron microscopy (LTEM) using Ga$^+$ focused-ion-beam (FIB) systems (FEI Nova 200 Dual-Beam SEM/FIB and Thermo Fisher Scientific Helios 5 CX DualBeam). The lamellae were measured to be $\sim 90$~nm thick for the $ab$-plane lamella (field along the $c$-axis) and $\sim 70$~nm thick for the $ac$-plane lamella (field along the $[120]$-axis).

\section{Lorentz transmission electron microscopy} LTEM measurements were performed on a JEOL NEOARM operated at 200~kV in low-magnification Lorentz mode with the objective lens turned off so that the specimen sits in a nominally field-free environment. Cryogenic imaging was carried out using a side-entry condenZero Javelin liquid helium flow cryostat, in which liquid helium is supplied from an external dewar through a low-noise transfer line, allowing the specimen to reach a nominal base temperature of 4.6~K as monitored by a Cernox sensor mounted on the holder 5~mm away from the sample~\cite{Kim2026Ultramicroscopy}. Prior to applying a magnetic field $B_{\text{ext}}$, which was generated by exciting the objective lens current, the lamella was oriented using selected area electron diffraction so that the field direction of interest lay approximately parallel to the optical axis, namely the $c$-axis for the $ab$-plane lamella and the $[120]$ direction for the $ac$-plane lamella. Images were recorded at a defocus of $300~\mu$m, which was re-established after each change in applied field.

\section{Computational Methods}
DFT calculations were performed using the Vienna \textit{ab initio} Simulation Package (VASP)~\cite{VASP-1,VASP-2,VASP-3}, including spin-orbit coupling (SOC). The projector augmented-wave (PAW) method~\cite{PAW} was employed, and the exchange--correlation energy was treated within the generalized gradient approximation (GGA) using the Perdew--Burke--Ernzerhof (PBE) functional~\cite{PBE}, with the corresponding PBE PAW potentials for Tm, Mn, Cr, and Sn. The experimental lattice parameters were kept fixed while internal atomic coordinates were relaxed. A plane-wave energy cutoff of $E_{\rm cutoff}=400$~eV and a $\Gamma$-centered $7\times7\times5$ $k$-point mesh were used. Spin-polarized calculations with SOC and constrained magnetic moments were used to compare different magnetic configurations and anisotropy energies. The \texttt{Tm\_3} PAW potential was used for Tm in the open-core approximation, where the localized Tm-$4f$ electrons are included in the core; when needed, a full-valence treatment of $f$ electrons was applied. This protocol avoids explicit treatment of the Tm-$4f$ manifold in the valence space while retaining contributions relevant to the Mn--Cr magnetic energetics.

\section{Decomposition of the Hall Resistivity}\label{sec:hall}

The Hall resistivity in a magnetic metal generally contains multiple contributions. The ordinary Hall term, arising from the Lorentz force on charge carriers, is present in all conductors and is proportional to the applied magnetic field. In systems with broken time-reversal symmetry, an additional contribution—the anomalous Hall effect (AHE)—appears, which is related to the magnetization and originates from both extrinsic mechanisms (skew scattering and side jump) and intrinsic Berry curvature of the electronic bands. In materials hosting non-coplanar spin textures with finite scalar spin chirality $\mathbf{S}_i \cdot(\mathbf{S}_j \times \mathbf{S}_k)$, where $\mathbf{S}_i$, $\mathbf{S}_j$, $\mathbf{S}_k$ are neighboring spins, an extra Hall response may arise from the real-space Berry phase acquired by conduction electrons, commonly referred to as the topological Hall effect (THE). Following the standard procedure employed in Ref.~\cite{noauthor_competing_nodate}, the total Hall resistivity can therefore be written as
\begin{equation}
    \rho_{xy} = \rho_0 + \rho_A + \rho_{xy}^{\mathrm{UHE}}.
    \label{Hall1}
\end{equation}

Here $\rho_0 = R_0B$ is the ordinary Hall resistivity, where $R_0$ is determined by the effective carrier density and mobility, $\rho_A = R_s\mu_0M$ is the conventional anomalous Hall resistivity, with $R_s$ the anomalous Hall coefficient and $M$ the magnetization, and $\rho_{xy}^{\mathrm{UHE}}$ denotes the residual unconventional (possibly topological) Hall (UHE) contribution [Eq.~(\ref{eqHall1}) of the main text].
To separate these terms, we assume that in the high-field forced-ferromagnetic regime the spin structure becomes collinear and any real-space Berry phase vanishes, such that $\rho_{xy}^{\mathrm{UHE}} \approx 0$. In this limit Eq.~(\ref{Hall1}) reduces to
\begin{equation}
\rho_{xy} = R_0B + R_s\mu_0M.   
\label{Hall2}
\end{equation}

The intercept of $\rho_{xy}/M$ vs.\ $B/M$ gives $\mu_0R_s$, while the slope gives $R_0$.

This procedure relies on several assumptions: (i) the anomalous Hall term scales linearly with magnetization over the fitting range, (ii) $R_0$ and $R_s$ are field-independent once the system is fully polarized, and (iii) multiband effects do not introduce significant curvature in the high-field Hall response. Deviations from these conditions can generate an apparent ``THE-like'' signal unrelated to real-space spin chirality, and this possibility is considered when interpreting the extracted $\rho_{xy}^{\mathrm{UHE}}$.

% \begin{figure}
%     \centering
%     \includegraphics[width=1\linewidth]{Supplementary/Fig01_Hall_M_MR_shared_axes.png}
%     \caption{Hall, magnetic, and magnetoresistive responses. (A–C) Hall resistivity \(\rho_{xy}\), (D–F) magnetization \(M\), and (G–I) magnetoresistance at 20, 40, and 50 K, respectively. Solid and dashed curves denote decreasing and increasing field, with arrows in (A–C) indicating the sweep directions. The 20 K panels show the low-field interval \(-2\leq B\leq2\) T, whereas the 40 and 50 K panels span approximately \(\pm9\) T. The comparison displays the low-field Hall branch separation alongside the magnetic and longitudinal responses, and the higher-field features at 40 and 50 K. Magnetization is expressed per formula unit.}
%     \label{All}
% \end{figure}

\section{Charge Transport and Magnetoresistance}\label{sec:mr}

\begin{figure}
    \centering
    \includegraphics[width=1\linewidth]{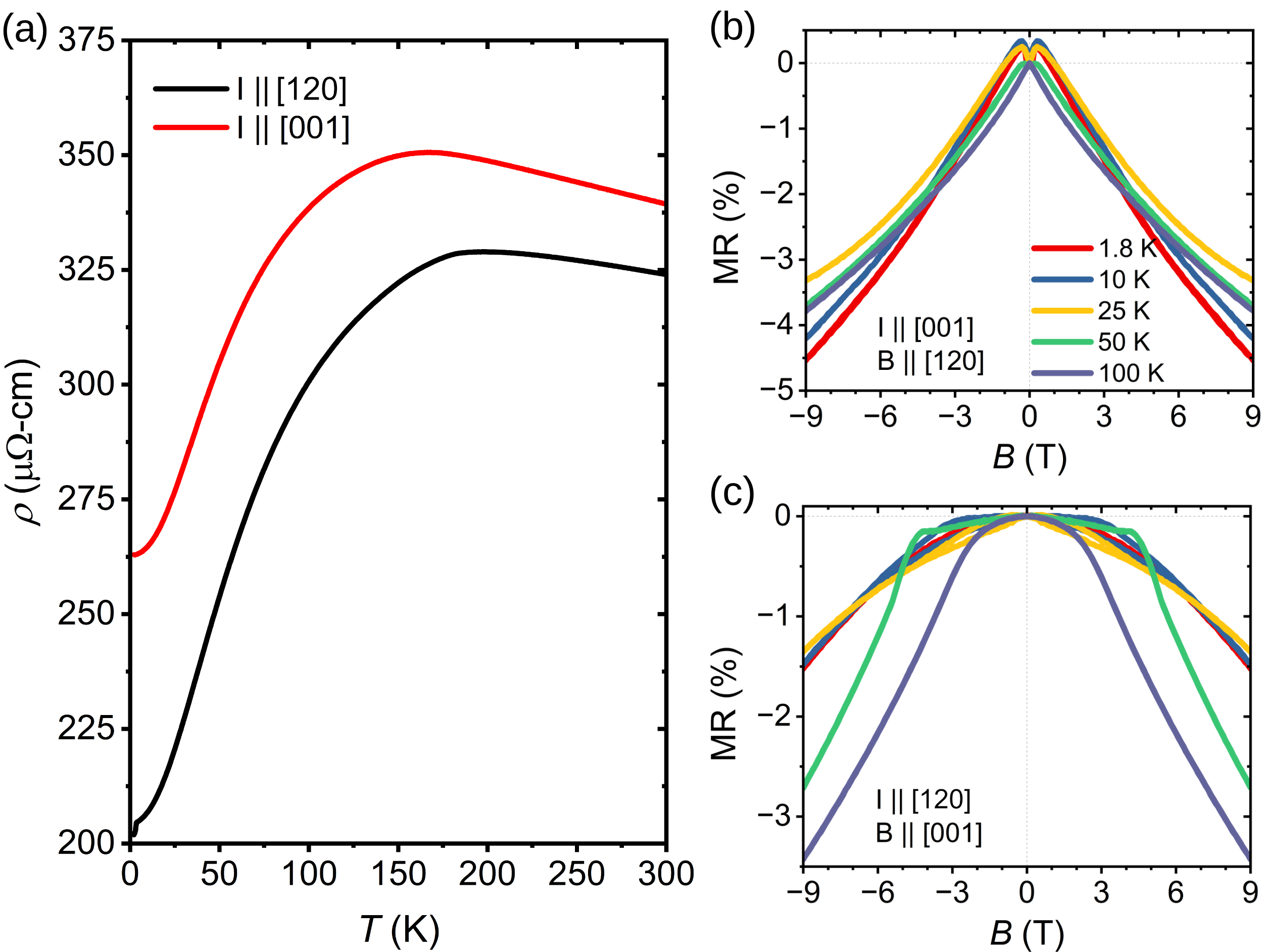}
    \caption{Charge transport in TmMn$_4$Cr$_2$Sn$_6$. (a) Longitudinal 
resistivity $\rho(T)$ for current $I \parallel [120]$ and 
$I \parallel [001]$. (b) Magnetoresistance MR(\%) for $B \parallel [120]$, 
$I \parallel [001]$ at selected temperatures. (c) Magnetoresistance for 
$B \parallel [001]$, $I \parallel [120]$ at selected temperatures.}
    \label{FigS3}
\end{figure}

The longitudinal resistivity $\rho_{xx}$ was measured with the current applied along both the $[001]$ (red) and $[120]$ (black) axes, as shown in Fig.~\ref{FigS3}(a). Upon cooling from $300$~K, both orientations exhibit a slight increase in resistivity down to approximately $200$~K, below which a metallic downturn develops. This crossover coincides with the onset of magnetic ordering observed in the susceptibility data (Fig.~\ref{Fig2}(a) of the main text), suggesting that magnetic ordering modifies the Fermi surface topology. The higher resistivity along $[001]$ compared to $[120]$ is opposite to the behavior observed in Ge- and Ga-doped $\mathrm{YMn_6Sn_6}$ \cite{bhandari_magnetism_2024}, but consistent with Cr-doped $\mathrm{TbMn_6Sn_6}$ \cite{destefano_giant_2025}, suggesting that transition-metal substitution on the Mn site alters the Fermi surface topology of $R\mathrm{Mn_6Sn_6}$ differently from Sn-site substitution.

Magnetoresistance (MR) was also measured in two geometries, $I \parallel [001]$, $B \parallel [120]$ (MR$_b$), and $I \parallel [120]$, $B \parallel [001]$ (MR$_c$), which are shown in Figs.~\ref{FigS3}(b) and \ref{FigS3}(c), respectively, for temperatures between $1.8$~K and $100$~K. Between $1.8$ and $25$~K, MR$_b$ increases up to $\sim 0.4$~T and then decreases. This crossover from increasing to decreasing MR is typically attributed to a change in scattering mechanism, be it the onset of a less resistive magnetic order, or simply a reduction in spin fluctuations. In this case, the saturation field along the [120]-axis sits around 0.3 T ($M$--$H$ curves in Fig.~\ref{Fig2}(d) of the main text), making this crossover from positive to negative MR consistent with this field-driven transition into an FM state. The transition into this FM state reduces MR by drastically lowering the scattering rates for one of the spin channels. At 50 K, the positive MR region is short-lived, and the MR almost immediately turns negative, while at 100 K, the MR is negative throughout the entire field range.  

MR$_c$ behaves very differently from MR$_b$. MR$_c$ is broad and negative across the entire applied field range. Between $1.8$~K and $25$~K, the data exhibit a clear hysteretic feature where the coercivity decreases with increasing temperature. Notably, while the corresponding magnetization ($M$--$H$) curves along the $[001]$ direction show only a very small hysteresis (Fig.~\ref{FigSHysteresis}), the MR curves in this same field orientation reveal significant hysteretic behavior. This discrepancy suggests that the electrical transport is highly sensitive to local magnetic scattering mechanisms that are not resolved by bulk macroscopic magnetization measurements.

At $50$~K, a distinct transition occurs at $4.1$~T, followed by a second, more subtle transition near $5.5$~T. This field range coincides with the field-driven spin reorientation observed in the $M$-$H$ curves along the $[001]$ axis, which saturate at roughly $5.8$~T. Following the interpretation in the main text, the sharper feature at $4.1$~T is associated with the discontinuous Mn/Tm spin reorientation, while the more gradual evolution up to $5.5$~T reflects the smooth canting of the weakly coupled Cr moments, which continues until the magnetization saturates. Beyond the $5.5$~T transition, the MR becomes linear, consistent with the onset of the saturated state. At $100$~K, the MR exhibits a broad quadratic downturn that gradually becomes linear near $3$~T without a distinct kink. This behavior is consistent with the magnetic saturation observed in the $M$-$H$ curves, which also smoothly occurs at approximately $3$~T.

\subsection{Raw transverse-signal comparison}

\begin{figure}
    \centering
    \includegraphics[width=1\linewidth]{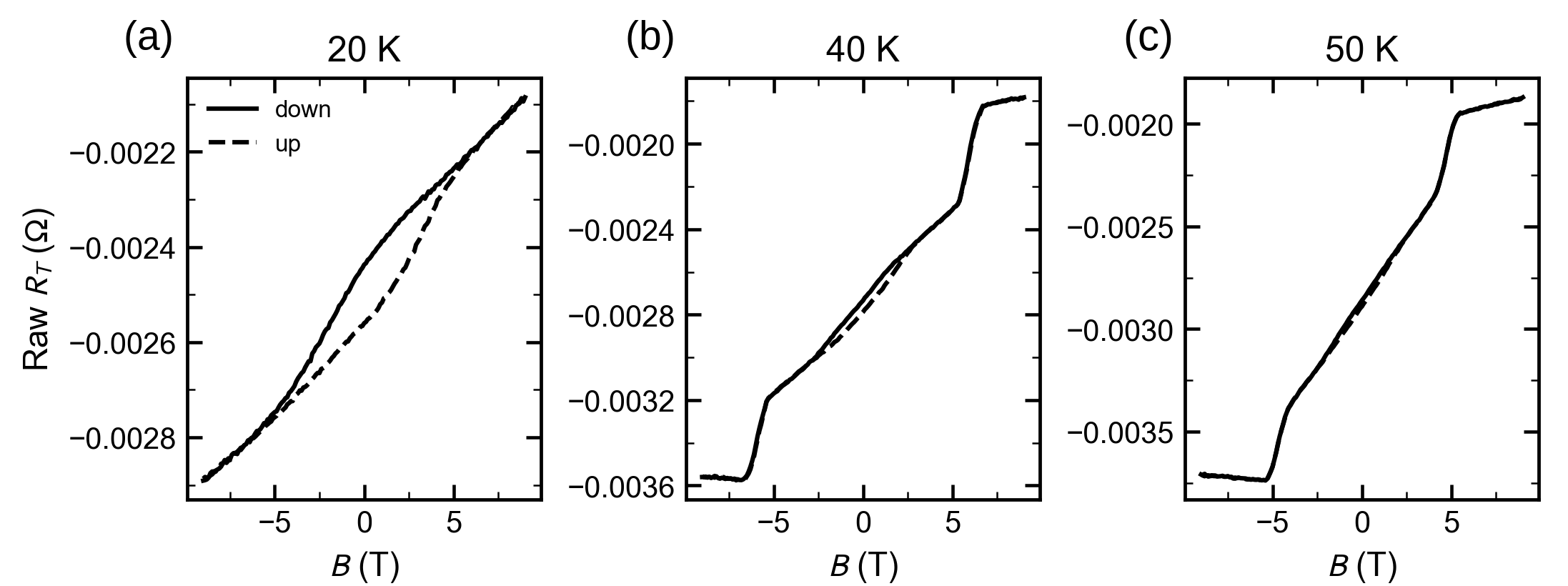}
    \caption{Raw transverse-resistance field sweeps. Transverse resistance \(R_T\) before antisymmetrization at (a) 20, (b) 40, and (c) 50 K. Solid and dashed curves denote decreasing and increasing field, respectively. The raw resistance offsets are retained. Low-field branch separation and the higher-field features at 40 and 50 K are visible in the raw transverse signal before the antisymmetrization was performed.}
    \label{Raw}
\end{figure}

Figure~\ref{Raw} shows the transverse resistance before antisymmetrization and conversion to resistivity. The separation between the two sweep histories is already visible in the raw 20~K trace, and the higher-field features at 40 and 50~K are also present before these processing steps. Their presence in the raw signal demonstrates that they were not generated solely by antisymmetrization or by the subsequent Hall-background fits. The absolute offsets of the raw curves are retained in this comparison and are not interpreted as Hall resistivity.

\subsection{Comparison of Hall resistivity, magnetization, and magnetoresistance}

\begin{figure}
    \centering
    \includegraphics[width=1\linewidth]{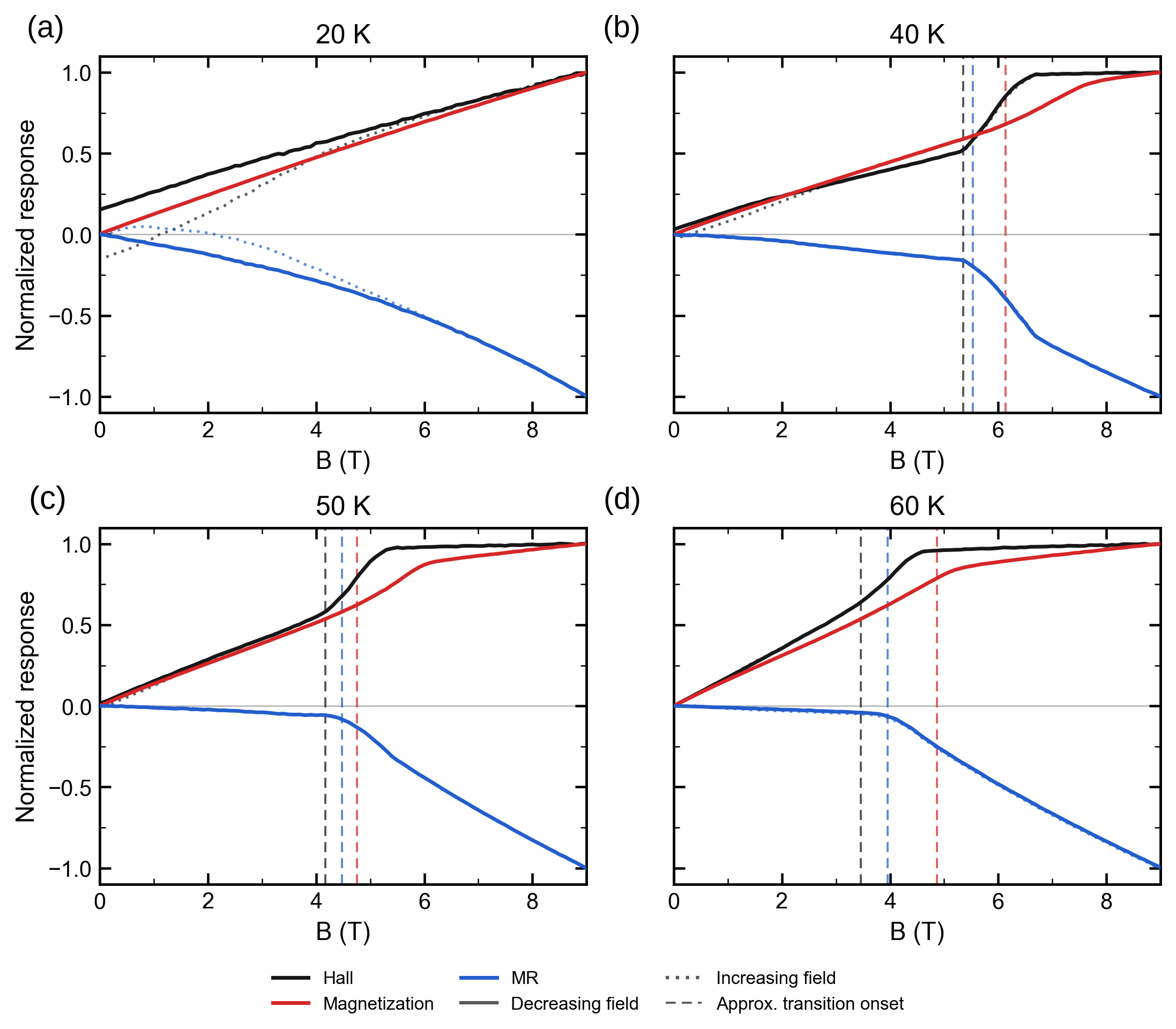}
    \caption{Comparison of the field dependence of Hall resistivity \(\rho_{xy}\) (black), magnetization \(M\) (red), and magnetoresistance (MR; blue) at (a) 20, (b) 40, (c) 50, and (d) 60 K. Each observable is normalized by its mean absolute value at \(+9\) T over the two sweep directions, retaining its measured sign. Solid and fine dotted curves denote decreasing and increasing field, respectively. At 40–60 K, three color-matched vertical dashed lines indicate approximate positive-field transition onsets.}
    \label{All}
\end{figure}

The normalized comparison in Fig.~\ref{All} shows how the field dependence of the Hall resistivity and MR responses compares with that of the magnetization. At 40 and 50~K, the pronounced increase in Hall resistivity and decrease in MR develop before the corresponding steepening of the magnetization is complete. The Hall signal subsequently approaches a much smaller high-field slope, whereas the magnetization and MR continue to evolve. This highlights the difference in apparent transition points between the Hall and magnetization measurements due to the canting of the Cr moments, leading to a remnant UHE.

\subsection{Low-Field Hysteresis Comparison}

\begin{figure}
    \centering
    \includegraphics[width=1\linewidth]{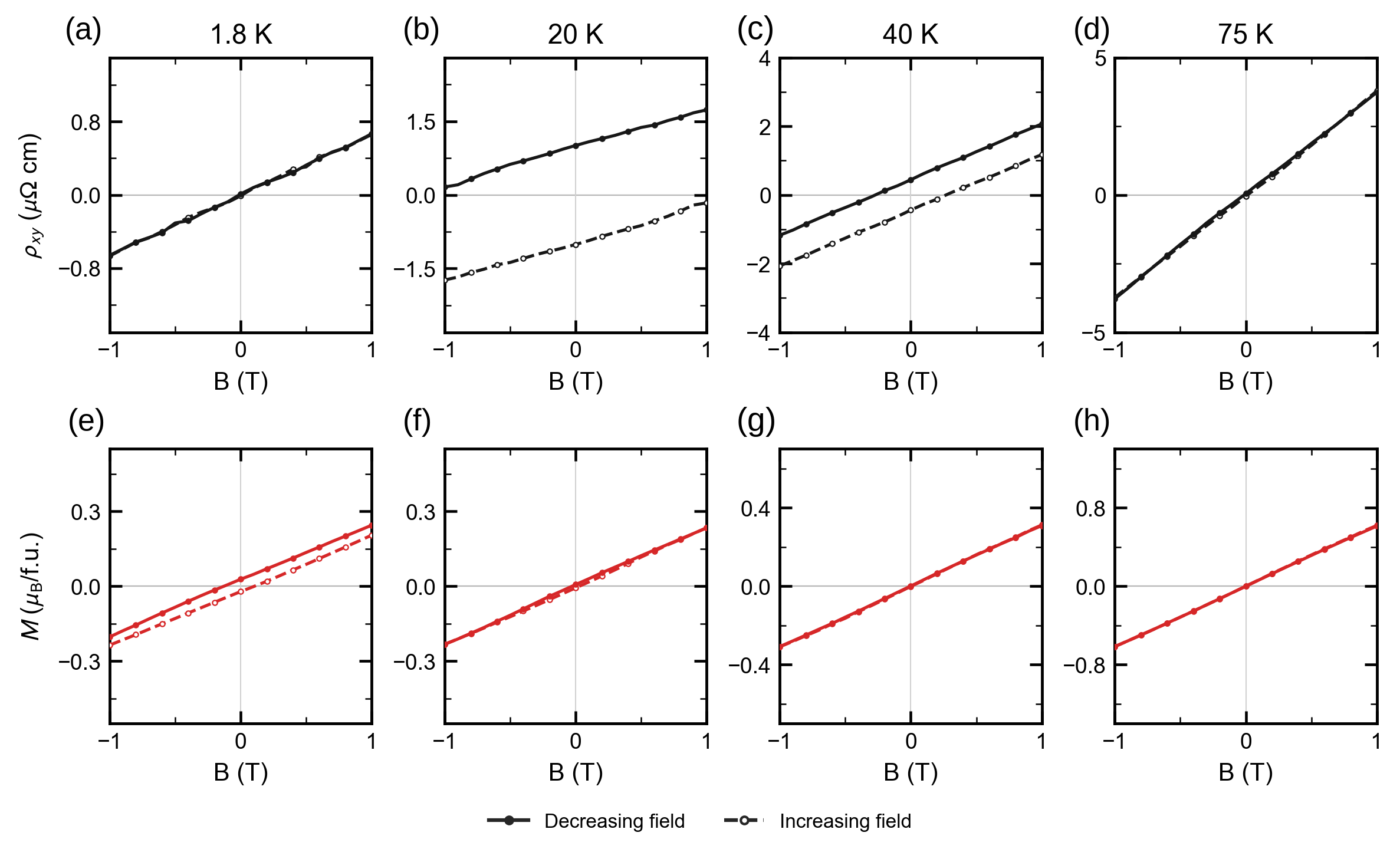}
    \caption{Low-field comparison of Hall and magnetization hysteresis. (a)--(d) Hall resistivity \(\rho_{xy}\) and (e)--(h) magnetization \(M\) at 1.8, 20, 40, and 75 K, respectively, over the common field interval \(-1\leq B\leq1\) T. Solid lines with filled symbols and dashed lines with open symbols represent decreasing and increasing field, respectively. Symbols are displayed at every second data point. Magnetization includes the original brass-background correction; no additional Hall-background subtraction or normalization is applied. Vertical scales vary between temperatures, as indicated by the tick labels.}
    \label{FigSHysteresis}
\end{figure}

Figure~\ref{FigSHysteresis} reveals contrasting temperature dependences of the low-field Hall and magnetization hysteresis. At 1.8~K, the magnetization branches are visibly separated, whereas the Hall branches nearly overlap. At 20~K, the Hall branch separation becomes pronounced while the magnetization separation decreases substantially. An appreciable Hall separation remains at 40~K, although the magnetization branches are nearly coincident on the displayed scale. At 75~K, both observables show comparatively small branch separations.

This contrast can be quantified using the zero-field branch half-difference,
\begin{equation}
X^{\mathrm{rem}}
=
\frac{X^{\mathrm{down}}(0)-X^{\mathrm{up}}(0)}{2},
\end{equation}

where $X$ denotes either $\rho_{xy}$ or $M$. Between 1.8 and 20~K, $\rho_{xy}^{\mathrm{rem}}$, which is obtained from the measured Hall branches before any background subtraction and therefore differs from the remanent $\rho_{xy}^{\mathrm{UHE}}$ quoted in the main text, increases from approximately $0.011$ to $1.009~\mu\Omega\,\mathrm{cm}$, whereas $M^{\mathrm{rem}}$ decreases from approximately $0.0248$ to $0.00663~\mu_{\mathrm{B}}$ per formula unit. At 40~K, the corresponding values are $0.441~\mu\Omega\,\mathrm{cm}$ and $0.00103~\mu_{\mathrm{B}}$ per formula unit. At 75~K, a small but discernible Hall hysteresis remains near zero field. The corresponding half-difference is $\rho_{xy}^{\mathrm{rem}} \approx 0.046~\mu\Omega\,\mathrm{cm}$. Thus, the Hall branch separation remains finite at 75~K even though the magnetization branches nearly overlap.

\subsection{Separation of Hall and magnetization derivative maxima}

\begin{figure}
    \centering
    \includegraphics[width=1\linewidth]{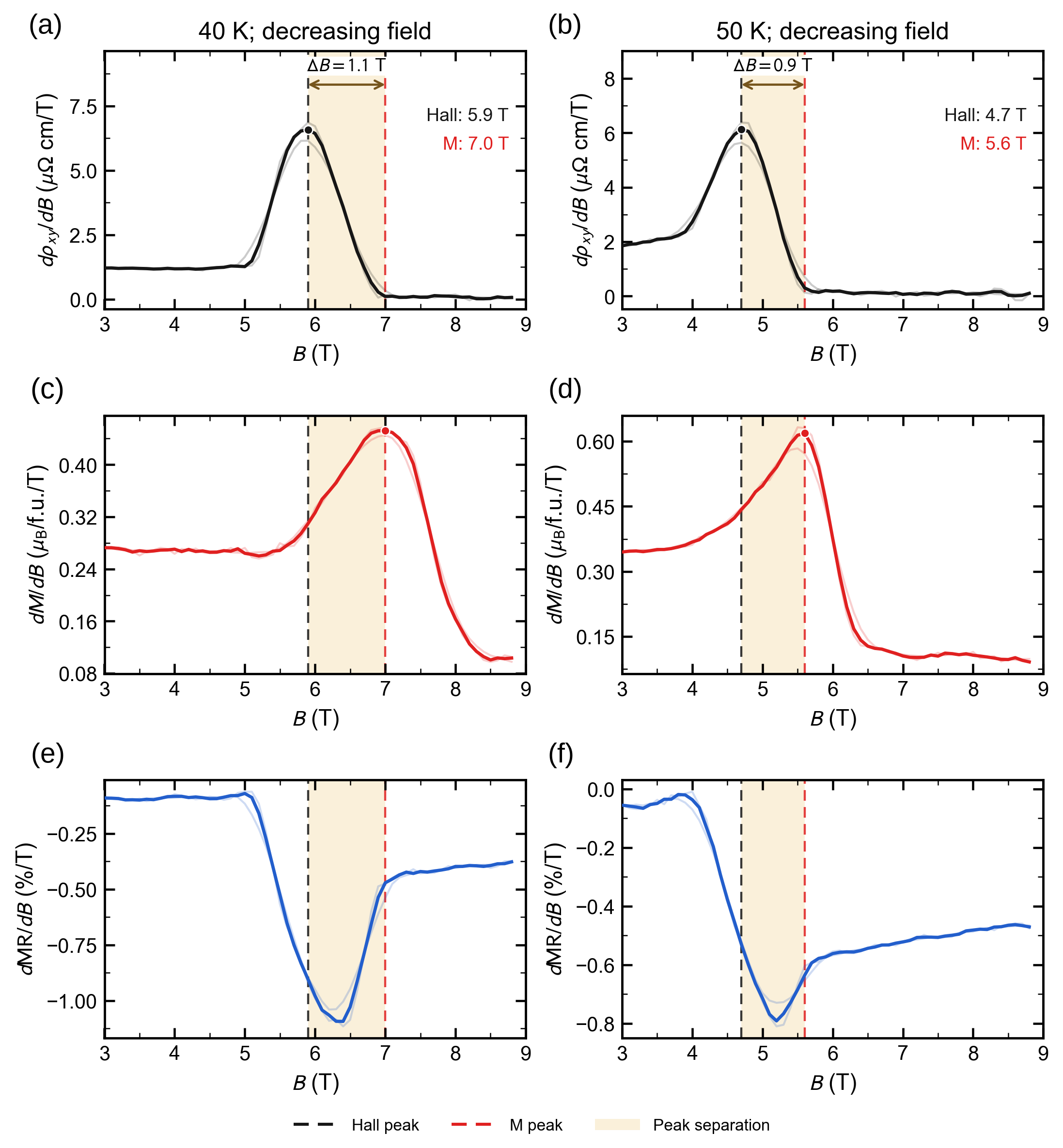}
    \caption{Comparison of high-field derivative features. Field derivatives of (a),(b) Hall resistivity, (c),(d) magnetization, and (e),(f) magnetoresistance on the decreasing-field sweep at 40 K (left column) and 50 K (right column). Bold curves were obtained using a quadratic Savitzky–Golay derivative with a seven-point window on a 0.1 T field grid; faint curves show five- and eleven-point windows. Circles and black/red vertical dashed lines identify the Hall/magnetization derivative maxima. These occur at 5.9/7.0 T at 40 K and 4.7/5.6 T at 50 K. Amber shading and arrows indicate the respective peak separations of 1.1 and 0.9 T. Shading represents peak separation, rather than transition width or uncertainty.}
    \label{Derivatives}
\end{figure}

The derivative comparison in Fig.~\ref{Derivatives} quantifies the different locations of the steepest Hall and magnetization responses. For the decreasing-field sweep at 40~K, $d\rho_{xy}/dB$ reaches its maximum near 5.9~T, whereas $dM/dB$ peaks near 7.0~T. At 50~K, the corresponding maxima occur near 4.7 and 5.6~T. The separations obtained with the seven-point derivative window are therefore approximately 1.1 and 0.9~T.

\subsection{Conductivity representation of the transport response}

\begin{figure}
    \centering
    \includegraphics[width=1\linewidth]{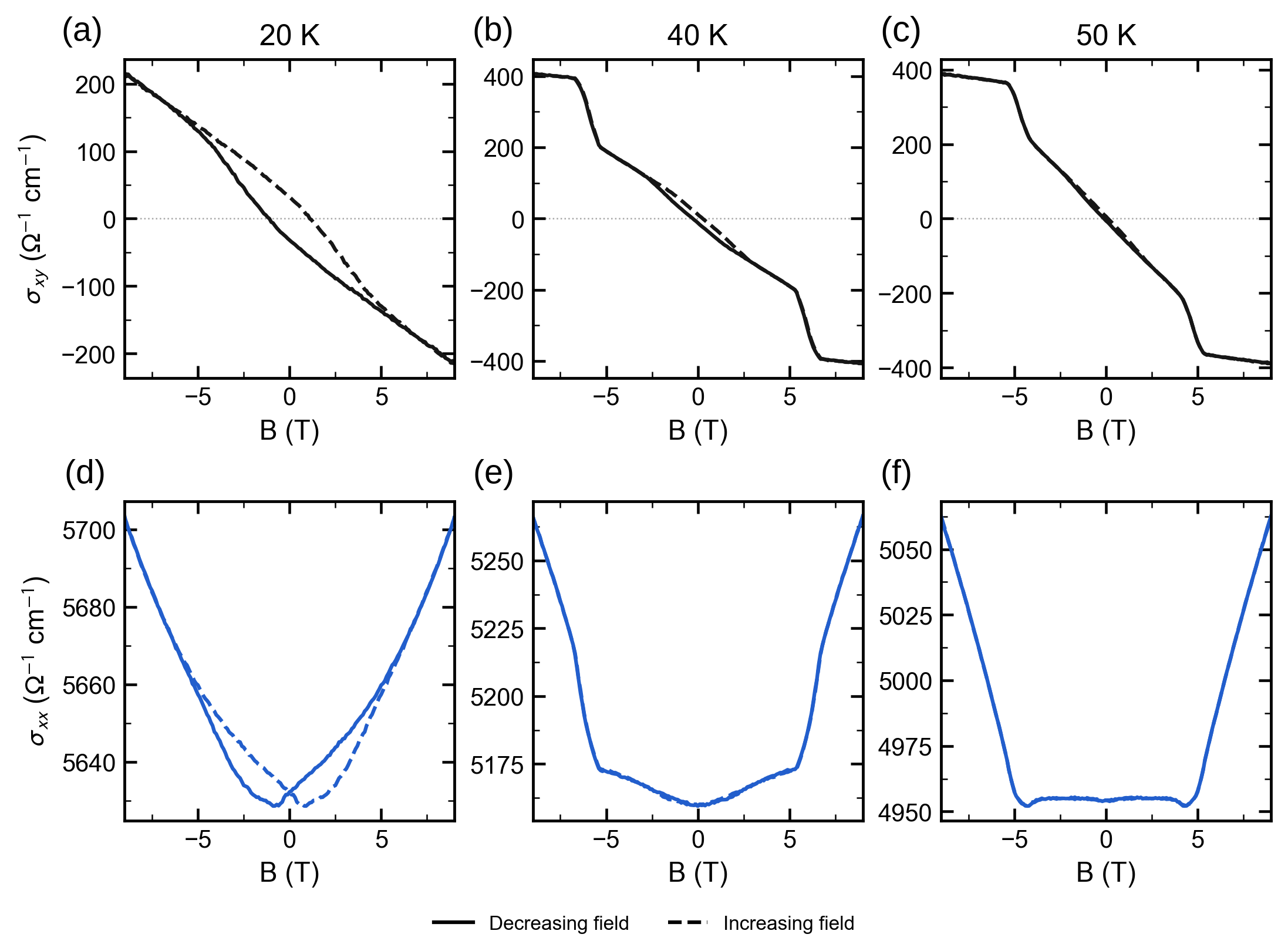}
    \caption{(a)--(c) Hall conductivity \(\sigma_{xy}\) and (d)--(f) longitudinal conductivity \(\sigma_{xx}\) at 20, 40, and 50 K, respectively. Solid and dashed curves represent decreasing and increasing field. Conductivities were calculated by pointwise inversion of the matched Hall and longitudinal resistivities, assuming \(\rho_{yy}=\rho_{xx}\), using \(\sigma_{xx}=\rho_{xx}/(\rho_{xx}^{2}+\rho_H^{2})\) and \(\sigma_{xy}=-\rho_H/(\rho_{xx}^{2}+\rho_H^{2})\). Here, \(\rho_H=E_y/J_x\) is the measured Hall resistivity, denoted \(\rho_{xy}\) in the resistivity figures. Both sweep histories are retained, showing the low-field hysteresis and higher-field changes in conductivity.}
    \label{Conductivity}
\end{figure}

Figure~\ref{Conductivity} examines whether the principal transport features remain apparent after conversion from resistivity to conductivity. The full tensor inversion was used, with

\[
D=\rho_{xx}^{2}+\rho_H^{2},\qquad
\sigma_{xx}=\frac{\rho_{xx}}{D},\qquad
\sigma_{xy}=-\frac{\rho_H}{D}.
\]

 Here, $\rho_H$ is the measured Hall resistivity. In the adopted off-diagonal convention, the resistivity tensor is $\left(\begin{smallmatrix}\rho_{xx}&\rho_H\\-\rho_H&\rho_{xx}\end{smallmatrix}\right)$, so that $\sigma_{xy}$ has the opposite sign to $\rho_H$. The inversion assumes equal diagonal resistivities in the transverse plane. Both sweep histories were retained in the calculation. Most importantly, the low-field Hall hysteresis at 20~K and the higher-field changes at 40 and 50~K remain visible in $\sigma_{xy}$. The longitudinal conductivity also changes through the high-field region, showing that the Hall feature occurs while the longitudinal transport is evolving.

\section{Real-Space Magnetic Imaging}\label{sec:ltem}

\begin{figure*}
    \centering
    \includegraphics[width=1\linewidth]{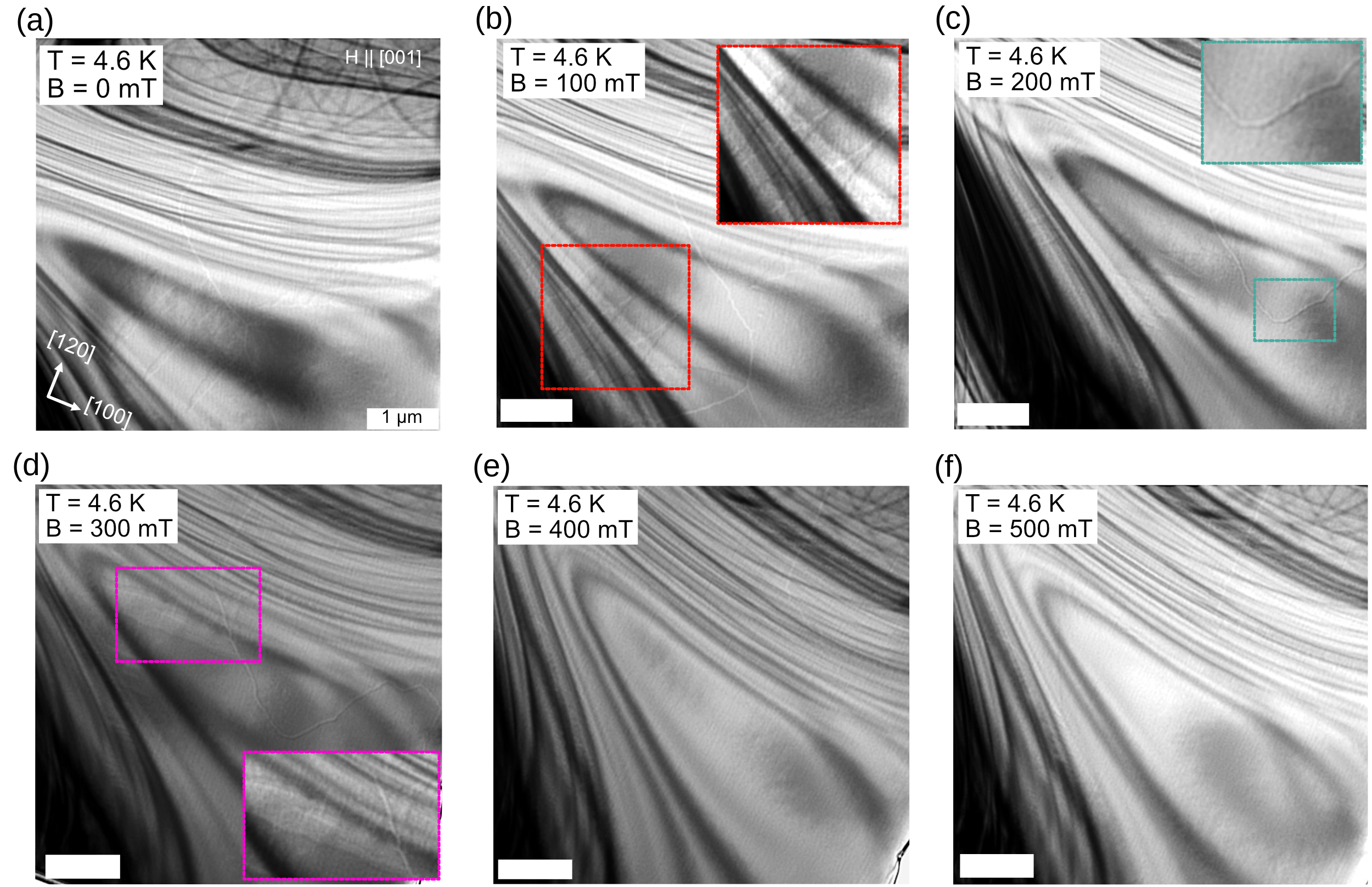}
    \caption{Lorentz transmission electron microscopy (LTEM) images of TmMn$_4$Cr$_2$Sn$_6$ at $T = 4.6$~K with $B \parallel [001]$ applied along the electron optical axis. (a)--(f) Evolution of the saw-tooth domain morphology under applied fields of 0, 100, 200, 300, 400, and 500~mT, respectively. Crystallographic directions $[120]$ and $[100]$ are indicated in (a). Scale bar: 1~$\mu$m. Dashed boxes indicate regions shown as enlarged insets.}
    \label{TEM_Bdep_c}
\end{figure*}

The LTEM imaging was performed in two distinct geometries. The first lamella was prepared in the $ab$-plane with the $c$-axis pointing out-of-plane, such that the applied field is parallel to the $c$-axis. This geometry replicates the field conditions under which the zero-field UHE was observed. The second lamella was prepared in the $ac$-plane with the [120]-axis oriented perpendicular to the lamella. Both samples were cooled to base temperature at 4.6~K in zero applied field, with the exception of the weak remanent field of the objective lens ($B_{\text{rem}} \approx 0.8$~mT).

Field-dependent LTEM images acquired on the first lamella at $T = 4.6$~K with $B \parallel [001]$ are shown in Figs.~\ref{TEM_Bdep_c}(a)--\ref{TEM_Bdep_c}(f). At zero field (Fig.~\ref{TEM_Bdep_c}(a)), we observe in-plane domains separated by a $180^\circ$ domain wall, which appears in the image as a bright-contrast line running through the middle of the lamella. The walls of this domain are characterized by sharp saw-tooth domains in the lower-left region of the lamella, oriented roughly along the $[\bar{2}\bar{1}0]$ direction. Upon application of 100~mT (Fig.~\ref{TEM_Bdep_c}(b)), these domains appear to reverse their orientation, now pointing along the [120] direction, while the bright-contrast domain wall remains largely unchanged.

At 200~mT (Fig.~\ref{TEM_Bdep_c}(c)), they rotate further, reorienting along the [100] axis, while the domain walls begin to deform. At 300~mT (Fig.~\ref{TEM_Bdep_c}(d)), a mixture of small and larger domain patterns emerges towards the left of the lamella, as highlighted in the pink inset. 
Saw-tooth domains remain resolvable but are smaller and more numerous. Above this field (Figs.~\ref{TEM_Bdep_c}(e)--\ref{TEM_Bdep_c}(f)), the saw-tooth domains can no longer be clearly resolved, and the bright-contrast domain wall progressively decays, remaining visible only in the upper region of the lamella, suggesting a transition to a larger domain with canted spins. We note that the sample is likely not polarized, because 500~mT is well below the saturation field measured in the bulk (Fig.~\ref{Fig4} of the main text).

\begin{figure*}
    \centering
    \includegraphics[width=1\linewidth]{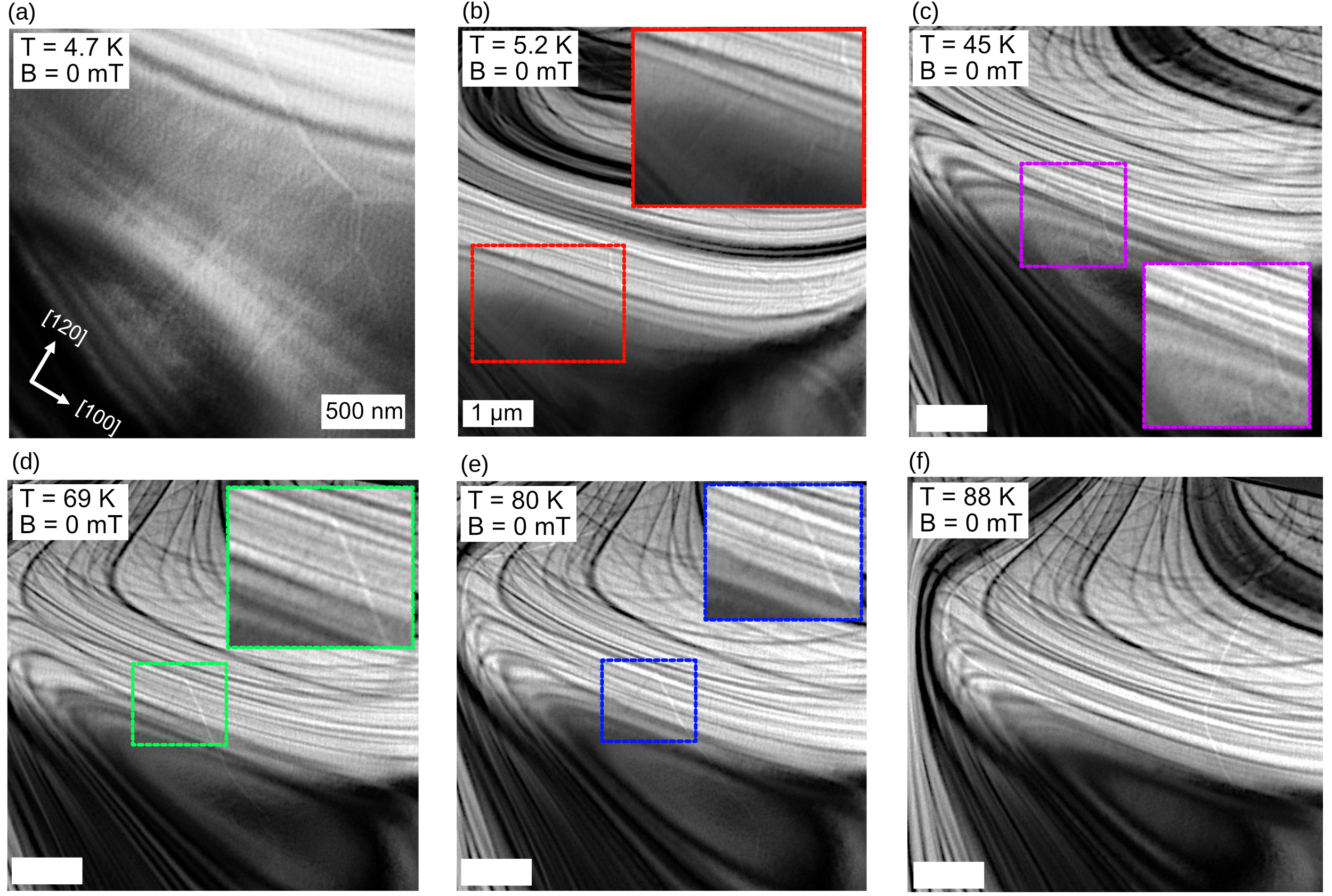}
    \caption{Zero-field LTEM temperature series within the $ab$ plane. (a)--(f) Domain structures at $T = 4.7$, 5.2, 45, 69, 80, and 88~K. The saw-tooth domains evolve markedly with temperature. Crystallographic directions $[120]$ and $[100]$ are indicated in (a). Scale bars: 500~nm (a), 1~$\mu$m (b)--(f).}
    \label{TEM_0B_ab}
\end{figure*}

\begin{figure*}
    \centering
    \includegraphics[width=1\linewidth]{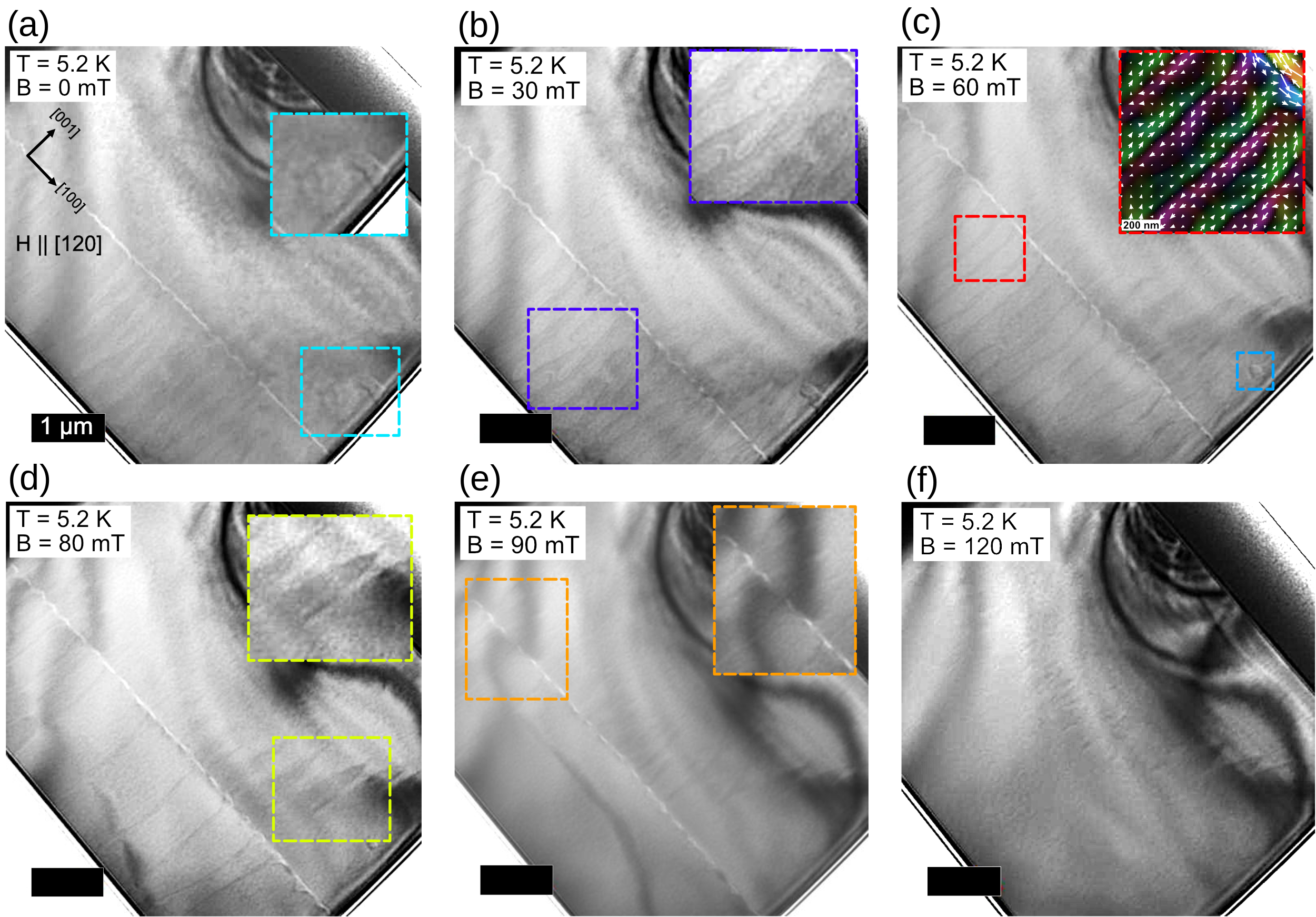}
    \caption{LTEM images at $T = 5.2$~K with $B \parallel [120]$ applied 
in-plane. (a)--(f) Domain evolution at applied fields of 0, 30, 60, 80, 
90, and 120~mT. Crystallographic directions $[001]$ and $[100]$ are 
indicated in (a). Inset in (c): single-image transport-of-intensity equation (SITIE) 
reconstruction of the in-plane magnetic induction over the region indicated 
by the dashed box, showing the magnetization texture at $B = 60$~mT. Scale 
bar: 200~nm (inset), 1~$\mu$m (main panels).}
    \label{TEM_Bdep_ab}
\end{figure*}

With zero magnetic field applied, the sample was warmed back to room temperature and the evolution of the magnetic domain structure was tracked as a function of temperature, as shown in Figs.~\ref{TEM_0B_ab}(a)--\ref{TEM_0B_ab}(f). At $T = 4.7$~K (Fig.~\ref{TEM_0B_ab}(a)), saw-tooth domains emanate from the in-plane domain wall. Upon a slight increase in temperature to $T = 5.2$~K (Fig.~\ref{TEM_0B_ab}(b)), these domains deform into a mixture of smaller saw-tooth domains and more unusual structures, as highlighted in the red inset. At $T = 45$~K (Fig.~\ref{TEM_0B_ab}(c)), the domain pattern bears a close resemblance to that observed at 5.2~K; however, the total number of domain structures present in the lamella is considerably reduced. The purple inset reveals a contrast feature that 
appears sharp at its upper edge but curves in an unusual fashion below.

At $T = 69$~K (Fig.~\ref{TEM_0B_ab}(d)), a clean, sharp saw-tooth domain structure re-emerges; however, it appears as a single domain much wider than those observed at low temperature. At $T = 80$~K and $88$~K (Figs.~\ref{TEM_0B_ab}(e)--\ref{TEM_0B_ab}(f)), this structure shows little further change.

\begin{figure*}
    \centering
    \includegraphics[width=1\linewidth]{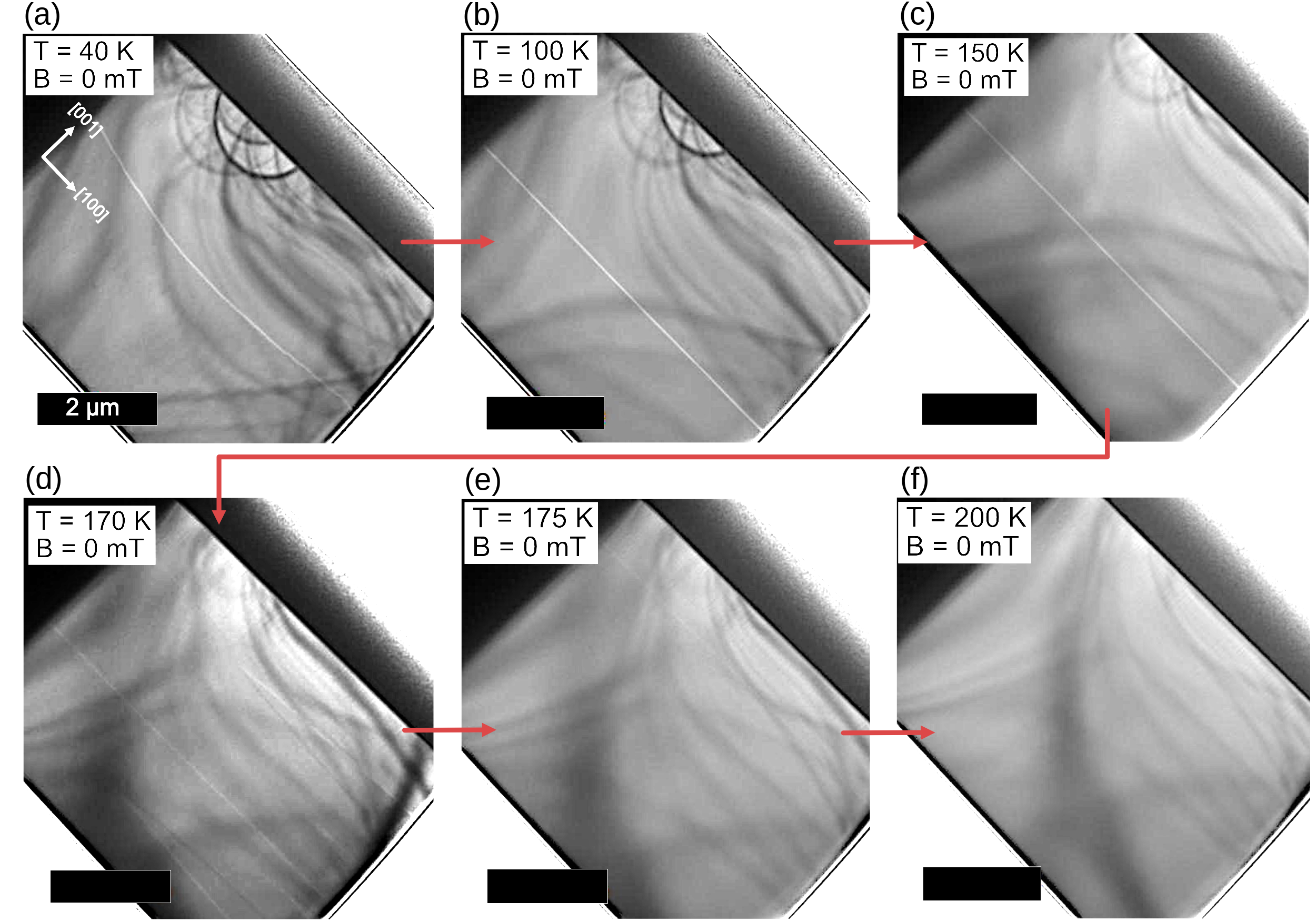}
    \caption{High-temperature LTEM image series at $B = 0$~mT within the $ac$ plane. (a)--(f) Domain structures at $T = 40$, 100, 150, 170, 175, and 200~K. Arrows indicate the sequence of domain wall evolution with increasing temperature. Crystallographic directions $[001]$ and $[100]$ are indicated in (a). Scale bar: 2~$\mu$m.}
    \label{TEM_0B_c}
\end{figure*}

We now present LTEM images acquired in the $ac$-plane geometry. Figure~\ref{TEM_Bdep_ab} shows the magnetic field dependence of the real-space magnetic textures with the field applied along the $[120]$-axis. The ZFC state at zero applied field (Fig.~\ref{TEM_Bdep_ab}(a)) reveals a striped texture in the lower half of the lamella, separated by a bright-contrast in-plane domain wall oriented along the $[100]$-axis. The stripes extend along the $[001]$-axis and are periodically spaced along the $[100]$ direction. Some contrast is visible in the right region of the lamella above the domain wall, but is strongly distinct 
from the striped pattern, as shown in the blue inset.

Upon application of 30~mT (Fig.~\ref{TEM_Bdep_ab}(b)), the stripes broaden slightly and additional magnetic contrast develops above the domain wall in the right region of the lamella. The striped domains exhibit a complex substructure, splitting into multiple regions with branching and merging contrast, as highlighted in the purple inset. At 60~mT (Fig.~\ref{TEM_Bdep_ab}(c)), the stripes are now well-defined as distinct, non-branching features. A SITIE reconstruction performed on the region indicated by the red box reveals an alternating spin arrangement, with moments pointing periodically along the $[001]$ and $[00\bar{1}]$ directions. The in-plane domain wall is also captured in the reconstruction, revealing a 180$^\circ$ rotation of the in-plane magnetization between the lower and upper regions of the lamella. Simultaneously, sharp saw-tooth domains of dark contrast 
appear in the upper region of the lamella, consistent with those observed in the $ab$-plane geometry discussed in Figs.~\ref{TEM_Bdep_c} and~\ref{TEM_0B_ab}. These saw-tooth domains emanate from the domain wall and point along the $[001]$-axis. We observe a small circular spin texture reminiscent of a type-II skyrmion bubble on the right side of the lamella with opposite-handed domain-wall contrast on its upper-right and lower-left sides, indicated by the dashed blue box.

At 80~mT (Fig.~\ref{TEM_Bdep_ab}(d)), the striped domains continue to broaden and new dark-contrast features begin to nucleate in the lower region of the lamella. These domains change shape but continue to occupy the upper region. At 90~mT (Fig.~\ref{TEM_Bdep_ab}(e)), the striped domains are fully suppressed and the lower region of the lamella is saturated along the $[120]$-axis. The saw-tooth domains remain visible but much fainter, and the in-plane domain wall is also weaker than at 80~mT. At 120~mT (Fig.~\ref{TEM_Bdep_ab}(f)), the in-plane domain wall is no longer resolvable, while the saw-tooth domains persist and appear to span the full width of the lamella.

After turning off the field, the sample was warmed back to room temperature and the evolution of the magnetic state at zero applied field was tracked, as shown in Figs.~\ref{TEM_0B_c}(a)--\ref{TEM_0B_c}(f). At 40~K (Fig.~\ref{TEM_0B_c}(a)), the in-plane domain wall remains bright and clearly visible; however, the striped domains observed at base temperature below this wall are much fainter and can barely be resolved. At 100~K and 150~K (Figs.~\ref{TEM_0B_c}(b)--\ref{TEM_0B_c}(c)), the in-plane domain wall persists and has shifted slightly toward the lower half of the lamella, though its contrast is noticeably reduced at 150~K relative to 100~K. At both temperatures, the striped domains are no longer visible.

At 170~K (Fig.~\ref{TEM_0B_c}(d)), two large saw-tooth domains appear, emanating from the right side of the lamella and pointing along the [100]-axis. Simultaneously, the in-plane domain wall becomes very weak in contrast, consistent with the system approaching its magnetic ordering temperature. Remarkably, just 5~K higher at 175~K (Fig.~\ref{TEM_0B_c}(e)), the saw-tooth domains disappear entirely and only a faint bright-contrast line is discernible in the upper-left region of the lamella, consistent with the same domain wall having shifted significantly. Finally, at 200~K (Fig.~\ref{TEM_0B_c}(f)), no magnetic contrast is resolvable and the sample is presumed to have entered the paramagnetic state.

The observed saw-tooth and stripe domains establish that Cr substitution produces nontrivial, field-sensitive domain structures, but their large size (hundreds of nanometers to microns) makes them highly unlikely to directly account for the magnitude of the observed UHE.

\end{document}